# Terradynamic streamlined shapes in animals and robots enhances traversability through densely cluttered, three-dimensional terrain

Chen Li[1,2,*], Andrew O. Pullin[2], Duncan W. Haldane[2], Han K. Lam[1], Ronald S. Fearing[2], and Robert J. Full[1]

[1]Department of Integrative Biology, University of California, Berkeley
5130 Valley Life Sciences Building, University of California, Berkeley, California, 94720-3140
[2]Department of Electrical Engineering & Computer Sciences, University of California, Berkeley
317 Cory Hall, University of California, Berkeley, California, 94720-1770
[*]Corresponding author at

## Abstract

Many animals, modern aircraft, and underwater vehicles use fusiform, streamlined body shapes that reduce fluid dynamic drag to achieve fast and effective locomotion in air and water. Similarly, numerous small terrestrial animals move through cluttered terrain where three-dimensional, multi-component obstacles like grass, shrubs, vines, and leaf litter also resist motion, but it is unknown whether their body shape plays a major role in traversal. Few ground vehicles or terrestrial robots have used body shape to more effectively traverse environments such as cluttered terrain. Here, we challenged forest-floor-dwelling discoid cockroaches (*Blaberus discoidalis*) possessing a thin, rounded body to traverse tall, narrowly spaced, vertical, grass-like compliant beams. Animals displayed high traversal performance (79 ± 12% probability and 3.4 ± 0.7 s time). Although we observed diverse obstacle traversal strategies, cockroaches primarily (48 ± 9 % probability) used a novel roll maneuver, a form of natural parkour, allowing them to rapidly traverse obstacle gaps narrower than half their body width (2.0 ± 0.5 s traversal time). Reduction of body roundness by addition of artificial shells nearly inhibited roll maneuvers and decreased traversal performance. Inspired by this discovery, we added a thin, rounded exoskeletal shell to a legged robot with a nearly cuboidal body, common to many existing terrestrial robots. Without adding sensory feedback or changing the open-loop control, the rounded shell enabled the robot to traverse beam obstacles with gaps narrower than shell width *via* body roll. Terradynamically "streamlined" shapes can reduce terrain resistance and enhance traversability by assisting effective body reorientation *via* distributed mechanical feedback. Our findings highlight the need to consider body shape to improve robot mobility in real-world terrain often filled with clutter, and to develop better locomotor-ground contact models to understand interaction with 3-D, multi-component terrain.

## 1. Introduction

To forage for food, find a mate, maneuver in a habitat, and escape predators, animals must move in environments that are spatially complex and temporally dynamic (Alexander 2003, Dickinson et al. 2000). By contrast, study of terrestrial locomotion began with much simpler environments. The first reduced-order mechanical and dynamical models of terrestrial locomotion were developed for animals walking, running, and hopping on flat, rigid, two-dimensional surfaces like treadmills and running tracks (Cavagna 1976, Blickhan 1989, Schmitt and Holmes 2000a,b) and climbing on flat, vertical walls (Goldman et al. 2006). More recently, we have begun to gain insights into terrestrial locomotion on ground with realistic topology, mechanics, and rheology such as uneven (Daley and Biewener 2007, Sponberg et al. 2008), sloped (Minetti et al 2002), low contact area (Spagna et al. 2007), low friction (Clark and Higham 2011), low stiffness (Spence et al. 2010, Ferris et al. 1998), damped (Moritz et al. 2003), and granular (Lejeune et al. 1998, Li et al. 2009) surfaces. While more representative than flat, rigid surfaces, most of these surfaces are still two-dimensional compared to the animal’s size and are relatively uniform.

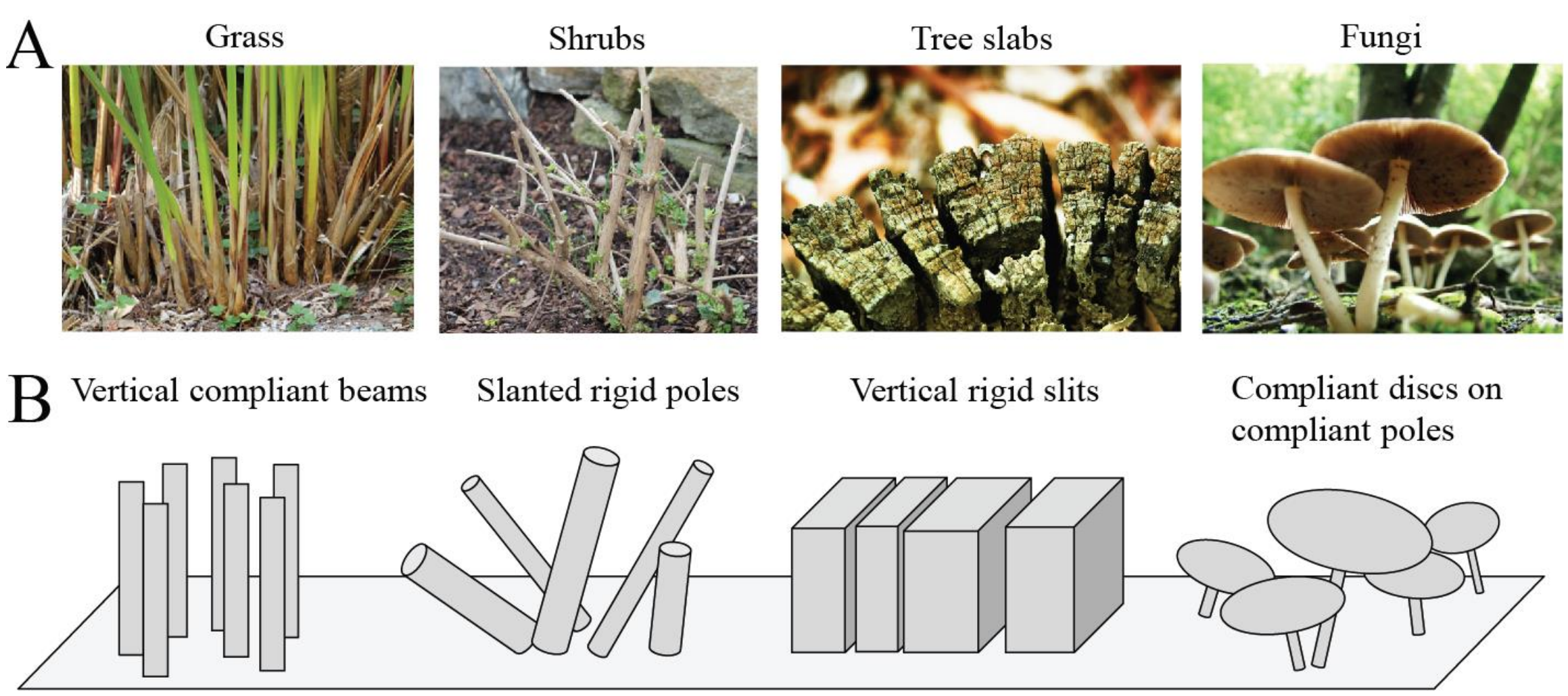

**Figure. 1.** Natural terrain is often filled with three-dimensional, multi-component obstacles. (A) Four representative types of 3-D, multi-component terrain and (B) their conceptual abstractions. [Photo credits: (B) Sheri Silver at sherisilver.com. (C) Craig Peihopa at Timeline Photography. (D)

This image entitled 'Giant 'shrooms' has been obtained by the author from the Flickr website where it was made available by wonderferret under a CC BY 2.0 licence.]

Here, we propose to advance terradynamics (Li et al. 2013) into three dimensions by going beyond relatively uniform, two-dimensional surfaces with three-dimensional obstacles of diverse, complex topology and mechanics, such as encountered in a forest floor with grass, shrubs, trees, and fungi (figure 1). In particular, small insects, arachnids, and reptiles face considerable challenges traversing such terrain, because these obstacles, which may be negligible for large animals, can be comparable or even much larger in size than themselves (Kaspari and Weiser 1999). Further, these obstacles can be densely cluttered with gaps, slits, and crevices comparable or even smaller than an animal's body, often pushing back against the animals, absorbing energy, and resisting locomotion, similar to surrounding fluids in flying and swimming. However, relative to fluids, the interaction of animals with such three-dimensional, multi-component terrain remains relatively unexplored (Jayne 1986, Summers & O'Reilly 1997, Aluck & Ward 2009, Tesch et al 2009, Qian et al 2013, Schiebel & Goldman 2014).

It is well known that animal shape can play an important role during locomotion in air and water (Vogel 1996, Lighthill 1960, Jacobs 1992; Betts & Wootton 1988, Swaddle & Lockwood 2003, Lovvorn & Liggins 2002, Norberg 1995). Almost all modern aerial and marine vehicles have adopted the fusiform, streamlined shapes common in large swimmers and fliers to reduce fluid-dynamic drag (Etkin 1972, Newman 1977). Similarly, for animals that must use their bodies to interact with and traverse three-dimensional, multi-component terrain, body shape might also affect locomotion. Yet, despite many ecological and functional morphology studies which established correlations between body and limb shapes and habitat use in insects, lizards, snakes, and birds (Burnham et al. 2010, Pike and Maitland 2003, Goodman et al. 2008, Herrel et al. 2002, Tulli et al. 2009, Fowler and Hall 2011, Navas et al .2004, Bickel and Losos 2002, Vanhooydonck & Van Damme 1999, Herrell et al 2001, Lovei & Sunderland 1996, Michael 1998, Thiele, 1977; Bell et al. 2007, Russell 1917, Forsythe 1983), few quantitative laboratory studies have investigated how body shape affects terrestrial locomotor behavior and biomechanics in complex terrain (Winter et al 2014, Sharpe et al, 2015).

Many insects, such as cockroaches (Bell et al. 2007) and beetles (Thiele 1977), have rounded body shapes with low angularity (see definition in Cho et al. 2006). Inspired by observations that rounded granular particles flow more easily through confined spaces such as orifices and hoppers (Li et al.

2004, Guo et al. 1985) and that thin, rounded objects like ellipses and ellipsoids are more difficult to grasp (Bowers and Lumia, 2003, Montana 1991, Howard and Kumar 1996), it is plausible that the rounded body shape of small insects may facilitate their rapid locomotion through cluttered obstacles. By contrast, most state-of-the-art wheeled (Iagnemma et al. 2008), tracked (Yamauchi 2004), and legged robots (Saranli et al 2001, Schroer et al 2004, Kim et al, 2006, Birkmeyer et al. 2009, Hoover et al. 2010, Pullin et al. 2012) have rectangular or cuboidal shapes with high angularity and low roundness. While much progress has been made in autonomous navigation of robots and vehicles (Thrun 2010) in sparsely cluttered terrain (obstacle spacing > robot size) using high level localization, mapping (Leonard & Durrant-Whyte 1991, Thrun et al. 2000) and motion/path planning (Latombe 1996), few existing ground vehicles and terrestrial robots have utilized effective body shapes to enhance traversability in cluttered terrain.

Inspired by streamlined shapes that reduce fluid dynamic drag, we hypothesized that a rounded body is terradynamically "streamlined" and can facilitate small animals' traversability in three-dimensional, multi-component terrain *via* effective mechanical interaction with cluttered obstacles. To test our hypothesis, we studied the locomotion of discoid cockroach (*Blaberus discoidalis*; figure 2A) through cluttered, grass-like beam obstacles with narrow spacing. Living on the floor of tropical rainforests, this animal frequently encounters and negotiates a wide variety of cluttered obstacles (see examples in figure 1). In addition, cockroaches are a good model organism because their mechanics and dynamics during rapid locomotion on simpler, near two-dimensional surfaces (Full & Tu 1990, 1991; Jindrich and Full 1999, 2000; Goldman et al. 2006, Spagna et al. 2007; Spence et al 2011) and their kinematics and sensory neural control during slow, quasi-static locomotion over three-dimensional terrain of simpler obstacles are well understood (Watson et al. 2002a,b, Harley et al 2009). Moreover, as opposed to slow locomotion that takes advantage of distributed neural feedback (Kindermann 2001, Roggendorf 2005, Harley et al. 2009, Ritzmann et al. 2005, Watson et al. 2002a,b), rapid locomotion in cockroaches primarily depends on distributed mechanical feedback *via* synergistic interaction of kinetic energy (Spagna et al 2007), morphology (Spagna et al 2007, Dudek et al 2006, Revzen et al. 2014), kinematics (Sponberg et al 2008), and material properties (Dudek & Full 2006) to resist perturbations.

Although a variety of studies have investigated small animals like insects and reptiles negotiating simple obstacles such as one or two steps, a gap, a ditch, an incline, or uneven surfaces (Watson et al. 2002a,b, Harley et al 2009, Blaesing and Cruse 2004, Theunissen and Durr 2013, Kohlsdorf and Biewener 2006; Byrnes and Jayne 2012, Ritzmann et al 2005, Bernd et al 2002, Theunissen et al.

2014), it remains a major challenge to parameterize and model the extraordinarily diverse and complex three-dimensional, multi-component obstacles found in nature (see examples in figure 1) so that controlled experiment can be conducted in the laboratory. Terrain parameterization by creation of controlled ground testbeds (Li et al 2009) has proven to be a powerful approach and facilitated the recent creation of the first terradynamic model that accurately and rapidly predicts legged locomotion for a diversity of morphologies and kinematics on a variety of granular media (Li et al. 2013).

Therefore, we take the next step in generalizing and advancing terradynamics (Li et al. 2013) by parameterizing three-dimensional, multi-component obstacles. We created a laboratory device enabling precise control and systematic variation of compliant vertical beam parameters (figure 3A). We used high-speed and standard imaging to record cockroaches traversing the beam obstacle field, and used an ethogram analysis (Harley et al 2009, Daltorio et al 2013, Blaesing and Cruse 2004) to quantify its locomotor pathway of traversal. To test our body shape hypothesis, we conducted direct experiments modifying the cockroach's body shape by systematically adding a series of artificial shells. We examined whether and how body shape change altered obstacle traversal pathways and performance. We used a physical model, a six-legged robot (figure 2B), to further test our hypothesis. Our animal discoveries provided biological inspiration for the robot to more effectively traverse grass-like beam obstacles. Finally, we developed a minimal potential energy landscape model to begin to reveal the importance of locomotor shape in multi-component terrain interactions.

## 2. Materials and Methods

### 2.1 Animals

For animal experiments, we used five adult male discoid cockroaches, *Blaberus discoidalis* (Mulberry Farms, Fallbrook, CA, USA), as females were often gravid and, therefore, under different load-bearing conditions. Prior to and during experimentation, we kept the cockroaches in communal plastic containers at room temperature (28°C) on a 12 h: 12 h light: dark cycle and provided water and food (fruit and dog chow) *ad libitum*. The discoid cockroach has a relatively thin, rounded body, the dorsal surface of which resembles a slice of an ellipsoid (figure 2A). See table 1 for animal body mass and dimensions.

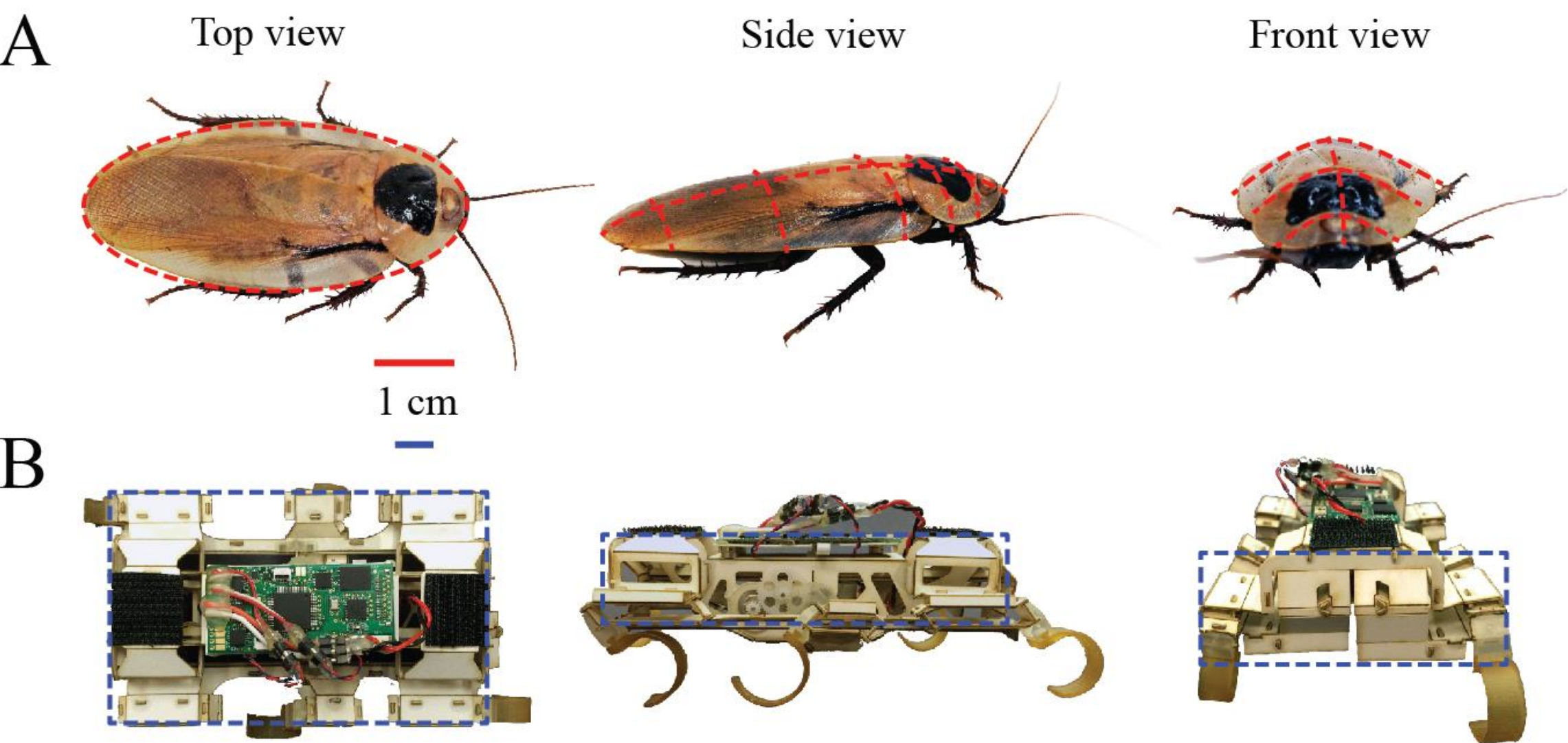

**Figure 2.** Model animal and robotic physical model. (A) The discoid cockroach has a thin, rounded body. (B) The VelociRoACH robot (Haldane et al. 2013) has a cuboidal body of high angularity and low roundness (see definition in Cho et al. 2006).

### 2.2 Legged robot

For robot experiments, we used VelociRoACH, a six-legged robot inspired by cockroaches (figure 2B, Haldane et al. 2013). The robot has springy c-legs and uses an alternating tripod gait to run at ~30 body length $s^{-1}$ with dynamics described by the Spring-Loaded Inverted Pendulum (SLIP) Model (Blickhan 1989, Full and Tu 1990). However, unlike the discoid cockroach, the robot has a cuboidal body, with flat faces, straight edges, and sharp corners (figure 2B), resulting in high angularity and low roundness. We ran the robot with an open-loop control algorithm to test only the effect of mechanical interaction. See table 1 for robot body mass and dimensions.

### 2.3 Beam obstacle track

We designed and constructed an apparatus to create a laboratory model of grass-like, cantilever beam obstacles that allowed precise control and systematic variation of the beam obstacles' geometric parameters and mechanical properties (figure 3A). The model beam track measured 180 cm long by 24 cm wide and was constructed using aluminum struts and acrylic (McMaster Carr, Elmhurst, IL, USA). The beams used in the track were manufactured by laser-cutting paper (VLS6.60, Universal Laser Systems, Scottsdale, AZ, USA). Control and variation of beam width, thickness, lateral spacing was attained by modifying laser cutting patterns (Adobe Illustrator,

Adobe Systems, San Jose, CA, USA). Control and variation of beam stiffness and damping was achieved by using different paper properties and modulation of beam dimensions. Control and variation of beam height, angle, and fore-aft spacing was attained by modification of the two parallel plates in which the beams were inserted and the two rails on which these two plates sat. For the animal experiments, we used thin cardstock (Wausau Exact Index Premium Cardstock, 199 g cm$^{-2}$, 279 × 216 mm$^2$, Wausau Paper, Harrodsburg, KY, USA) to manufacture the beams. For the robot experiments, we used thick cardstock (Pacon 6-ply railroad board, Appleton, WI, USA). Mechanical tests showed that the model beams had a Young's modulus (~10$^9$ Pa) within the natural range that we found for small grass and thin plant stems (~10$^8$ – 10$^{10}$ Pa) by direct measurement, and thus a similar stiffness due to their similar geometry.

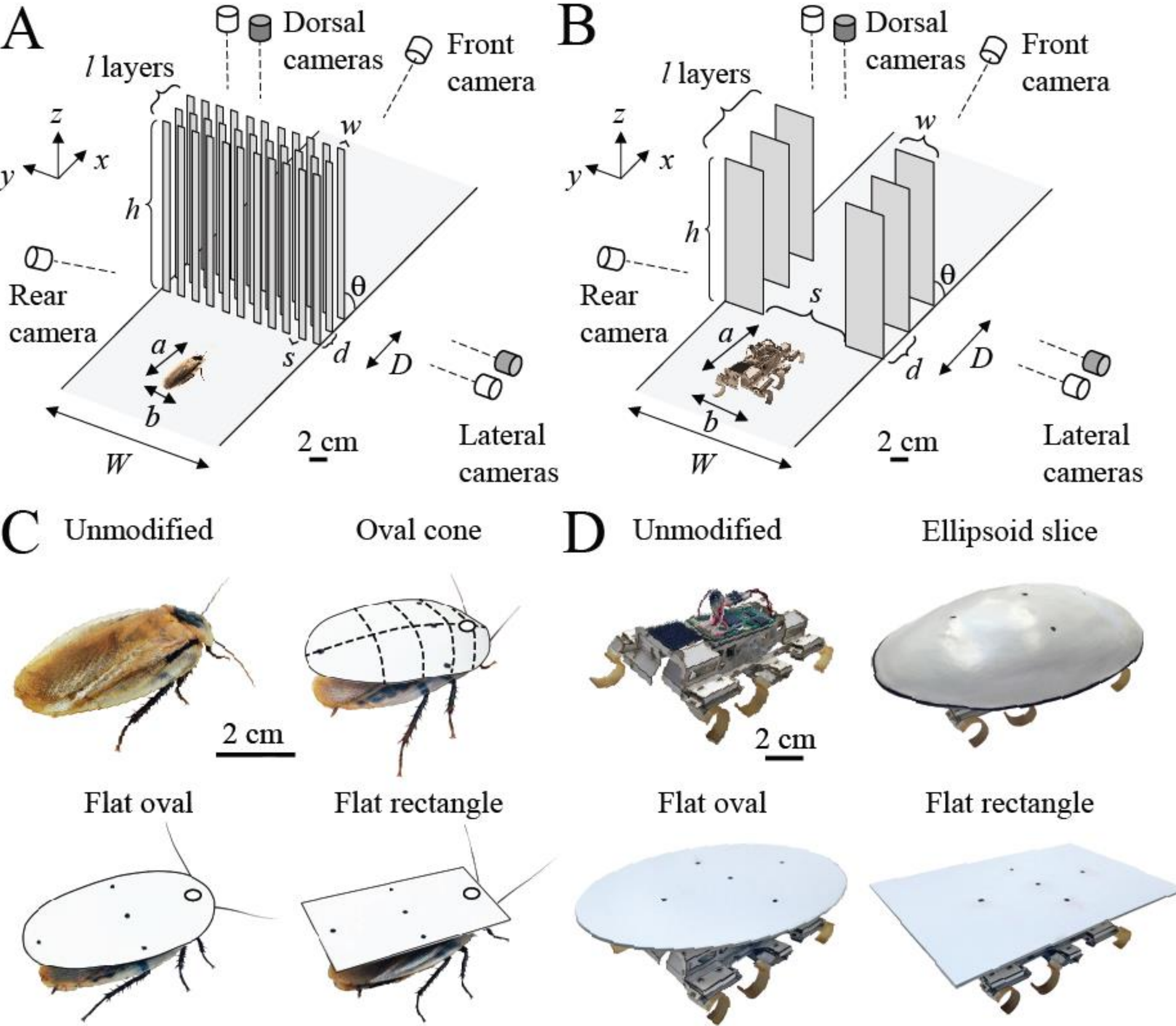

**Figure 3.** Experimental setup and body shape modification. Beam obstacle field and imaging setup in animal (A) and robot experiments (B). The apparatus allowed control and variation of beam geometric parameters and mechanical properties including number of layers ($l$), width ($w$),

thickness, lateral spacing ($s$), fore-aft spacing ($d$), height ($h$), and angle ($\theta$), stiffness, and damping. White: regular-speed cameras; gray: high-speed cameras. Modification of the animal's (C) and robot's (D) body shape by adding shells of three different shapes.

Table 1. Body mass, dimensions, and beam obstacle parameters for animal and robot experiments.

| Parameter | Animal experiments | Robot experiments |
|---|---|---|
| Body mass $m$ (g) (unmodified) | 2.5 ± 0.3 | 26 |
| Body length $a$ (cm) | 4.9 ± 0.1 | 10 |
| Body width $b$ (cm) | 2.4 ± 0.1 | 6.6 |
| Body thickness $c$ (cm) | 0.7 ± 0.0 | 3 |
| Standing height $c'$ (cm) | 1.2 ± 0.1 | 4 |
| Beam layers number $l$ | 3 | 3 |
| Beam angle $\theta$ (°) | 90 | 90 |
| Beam height $h$ (cm) | 10 | 27 |
| Beam width $w$(cm) | 1 | 5.5 |
| Beam lateral spacing $s$ (cm) | 1 | 10 |
| Beam fore-aft spacing $d$ (cm) | 2 | 4 |
| Beam obstacle field width $W$ (cm) | 21 | 21 |
| Beam obstacle field depth $D$ (cm) | 4 | 8 |

All average data are given as mean ± s.d., unless otherwise specified.

In animal experiments, we challenged cockroaches to go through a beam obstacle field (figure 3A) consisting of three layers of vertical, tall beams taller than twice the animal body length. Lateral spacing was less than half the animal's body width, slightly less than the animal's standing height, but slightly larger than the animal's body thickness without legs. The fore-aft spacing between each adjacent layer was less than half the animal's body length (table 1).

In robot experiments, the beam obstacle field (figure 3B) consisted of three layers of vertical, tall beams, Beams were taller than twice the robot body length, with lateral spacing larger than the robot's body width, but smaller than the width of the artificial shells (12 cm, figure 3D; see Section 2.4). The fore-aft spacing between each adjacent layer was less than half the robot body length (table 1).

### 2.4. Shape modification experiments

We first tested the animal with an unmodified body shape. To determine whether and how an animal's body shape affected its traversability through beam obstacles, we further tested the same individuals with modified body shapes by sequentially adding artificial shells to the animal's dorsal surface with the following shapes (figure 3C, table 2): (1) an oval cone with similar rounded shape to the animal's body; (2) a flat oval to remove three-dimensional roundness; and (3) a flat rectangle to further remove two-dimensional roundness. Finally, to control for long-term learning and fatigue, we tested unmodified animals again after removing the shells. In preliminary experiments, we randomized the order of presentation of the three shells and found no effect. By testing all shells on the same individuals, they served as their own control.

Table 2. Animal mass, shell mass, and number of trials for animal experiments.

| Animal ± shell | Animal mass (g) | Shell mass (g) | Number of trials (*n*) |
|---|---|---|---|
| No shell (before) | 2.6 ± 0.3 | N/A | 176 |
| Oval cone shell | 2.5 ± 0.2 | 0.4 ± 0.0 | 170 |
| Flat oval shell | 2.5 ± 0.3 | 0.4 ± 0.1 | 168 |
| Flat rectangle shell | 2.5 ± 0.3 | 0.5 ± 0.0 | 204 |
| No shell (after) | 2.5 ± 0.2 | N/A | 151 |

All average data are given as mean ± s.d., unless otherwise specified. $N = 5$ individuals.

We used a small amount of hot glue to securely attach the artificial shells to the animal's hard pronotum. The shells were made of thick cardstock (Pacon 6-ply railroad board, Appleton, WI, USA). The maximal length and width of the three shells for each animal were the same as its body length and width. The shells represented a small increase in the animal's mass (table 2), but significantly increased the overall body volume and surface friction. The kinetic friction coefficient was 0.54 ± 0.03 (measured by the inclined plane method) between the shell and beams, greater than 0.10 ± 0.01 between the animal body and beams, ($P < 0.05$, Student's $t$-test). The animal running on flat ground did not slow down with the shells (64 ± 11 cm $s^{-1}$) compared to without shells (67 ± 11 cm $s^{-1}$; $P > 0.05$, repeated-measures ANOVA).

Table 3. Weight, dimensions, and manufacturing of robot exoskeletal shells.

| Shell | Mass (g) | Length (cm) | Width (cm) | Height (cm) | Manufacturing | Material |
|---|---|---|---|---|---|---|
| Ellipsoidal shell | 19 | 18 | 12 | 3 | thermoforming | 0.75 mm polystyrene |
| Flat oval shell | 21 | 18 | 12 | N/A | laser cutting | 1.5 mm polystyrene |
| Flat rectangle shell | 26 | 18 | 12 | N/A | laser cutting | 1.5 mm polystyrene |

From our hypothesis that the discoid cockroach's thin, rounded body facilitates its ability to traverse three-dimensional, multi-component terrain, we predicted that: (1) with unmodified body shape, the animal's obstacle traversal performance will be at its maximum; (2) with the oval cone shell, traversal performance will decrease due to increased volume and surface friction; and (3) with the flat oval and rectangle shells performance will be further decreased as roundness is reduced, and obstacle traversal pathways altered. Assuming negligible learning and fatigue, we expected the animal to recover its obstacle traversal pathways and performance after the shells were removed.

In robot experiments, we also modified the robot's body shape by adding exoskeletal shells of similar shapes to those in the animal experiments (figure 3D): an ellipsoidal shell (Haldane et al 2015), and a flat oval shell, and a flat rectangle shell (table 3). Although all the shells were comparable to the robot in mass, adding them did not slow down the robot on flat ground (60 cm $s^{-1}$ with or without shells at 10 Hz).

### 2.4.1 Experimental protocol

We used four webcams (C920, Logitech, Newark, CA, USA) to record entire experimental sessions from top, side, front, and rear views at 30 frame $s^{-1}$ for analysis of the animal and robot's obstacle traversal pathways (figure 3A). A custom four-way LED array sent signals to synchronize the four webcams. Two high-speed video cameras (Fastec) were set up to record simultaneous top and side views with 1280 × 540 pixel resolution at 250 frame $s^{-1}$ to capture detailed kinematics of representative trials. We adjusted the side high-speed camera's lens to a small aperture size to maximize depth of field such that the entire width (21 cm) of the beam obstacle field was in focus. An external trigger synchronized the two high-speed cameras. Eight 500 W work lights illuminated the experimental area from the top and side to provide ample lighting for the high-speed cameras.

In animal experiments, we set up two walls before the beam obstacle field to funnel the animal's approach toward the middle of the obstacle field, far away from the sidewalls of the track. We placed a piece of egg carton after the obstacle field to encourage the animal to traverse the obstacle field and seek shelter (Harley et al. 2009; Daltorio et al. 2013). Before each experimental session, the test area was heated to 35°C by the work lights. The track was illuminated during the entire experimental session to maintain temperature. During each trial, we released the cockroach onto the approach side of the track, and elicited a rapid running escape response by gently probing the posterior abdominal segments and cerci with a small rod wrapped with tape. The animal quickly ran down the track through the funnel, and attempted to traverse the beam obstacle field. If the animal succeeded in traversing the field it would then enter the egg carton shelter. If the animal failed to traverse within 40 seconds, it was picked up and placed into the shelter. The animal was then allowed to rest at least one minute within the shelter before the next trial.

In robot experiments, at the beginning of each trial, we set the robot at 18 cm from the center of the robot to the first layer of two beams and in the middle of the two beams, and carefully positioned the legs in the same phase for all trials. During each trial, we ran the robot at 10 Hz stride frequency for 15 seconds. We swapped the battery every three trials to ensure that the voltage remained nearly constant throughout all trials.

#### 2.4.2 Data analysis and statistics

From the videos, we obtained the animal and robot locomotor performance metrics including traversal probability and time. We also obtained the locomotor ethogram for each trial, and performed a locomotor pathway ethogram analysis (Harley et al 2009, Daltorio et al 2013, Blaesing and Cruse 2004) to quantify the locomotor pathways of traversal.

In animal experiments, we collected a total of 1,050 trials from five individuals each with five shape treatments, with approximately 40 trials for each individual and shape combination. We rejected trials in which the animal spontaneously ran towards the beam obstacle field (rather than after being chased by the experimenter), because during spontaneous locomotion the animal often slowly explored and traversed the obstacle field rather than attempting to negotiate it as fast as it could during the escape response. Including trials with slow exploration would bias the traversal time results. We also rejected trials in which the animal used the sidewalls of the track to negotiate the beam obstacles. With these criteria, we accepted 869 trials from five individuals each with five shapes, approximately 35 trials for each individual and shape combination. This large sample size enabled us to obtain statistically meaningful data of the distribution and pathways of the diverse locomotor modes observed (see Section 3.1).

To analyze obstacle traversal pathways and performance for each individual and shape treatment, we calculated the averages of all locomotor metrics (e.g., traversal probability, traversal time, transition probability between locomotor modes) using all the trials from the same individual with each shape treatment. To compare between each shape treatment, we then calculated the averages of each locomotor metric by averaging the means of all five individuals for each shape treatment. All average data are given in mean ± s.d., unless otherwise specified. We used repeated-measures

ANOVA to test whether each locomotor metric (averages of the means of five individuals) differed among body shapes. We used Tukey's honestly significant difference (HSD) test for post-hoc analysis where needed.

To begin to differentiate the effects of body shape *versus* active feedback resulting in altered leg thrusts, we pulled deceased animals through the vertical beam field (figure 4). A stepper motor (ROB-09238, SparkFun Electronics, Boulder, CO, USA) pulled freshly deceased animals through the beam obstacle field using a wire with an end glued to the dorsal surface of the animal's pronotum at a constant speed (10 cm $s^{-1}$), while videos were recorded from top, side, and rear views using three webcams (C920, Logitech, Newark, CA, USA). Videos were analyzed to determine the movement of the animal body during interaction with the beams. We added the same three shell shapes to the deceased animals. For each shape treatment, we performed ten pulling trials. At the beginning of each pulling trial, the animal body was placed at the same location, and care was taken to ensure that the wire was in the middle of two adjacent beams with minimal friction.

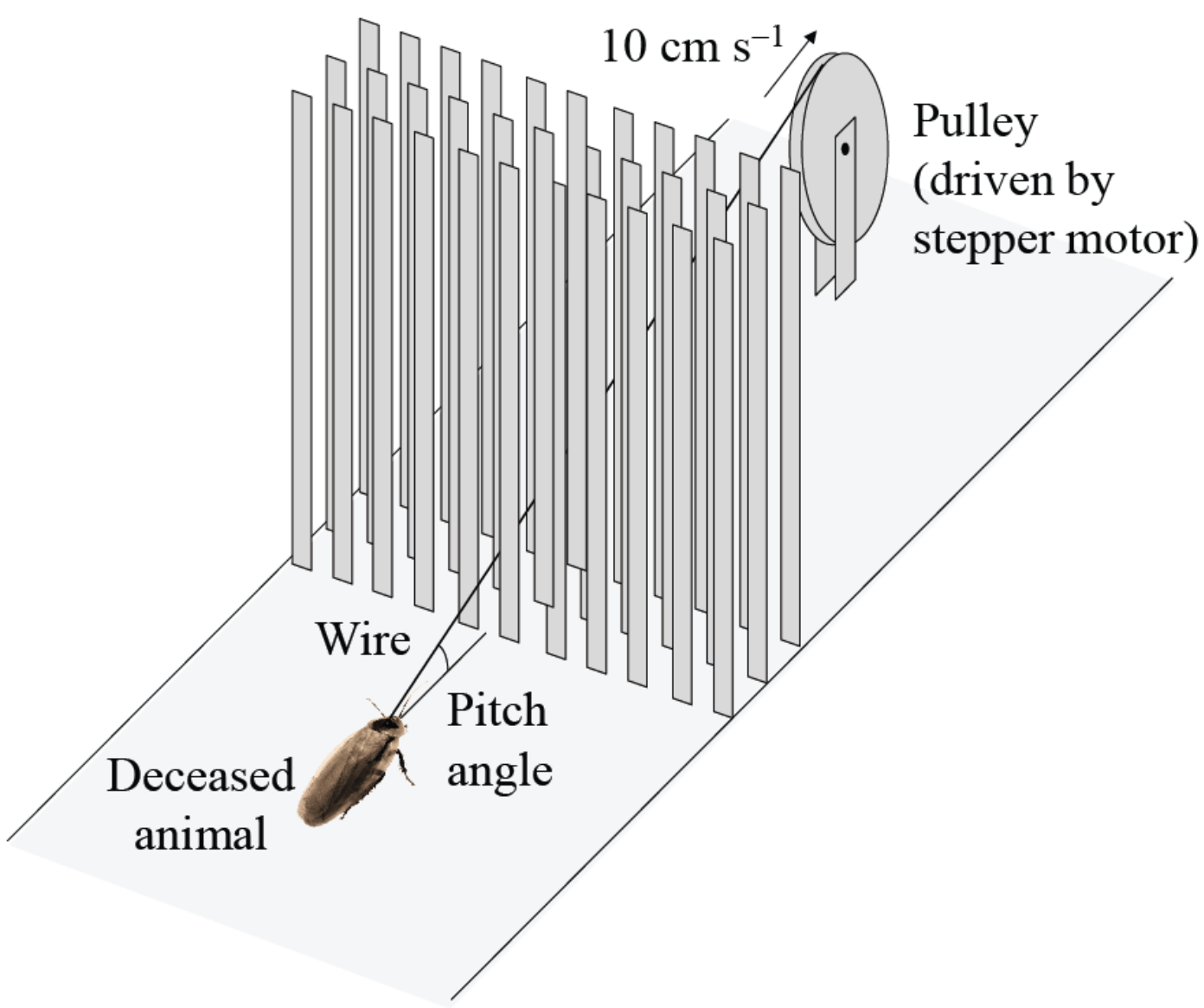

**Figure 4**. Experimental setup to pull deceased animals through beam obstacles using a stepper motor. We also did this using the same shells attached to the live running animals.

In robot experiments, we collected 15 trials for each shape treatment, and used ANOVA for statistical testing.

## 3. Results and Discussion

### 3.1 Diverse locomotor pathways to traverse obstacles

We first focused on the animal's beam obstacle traversal with an unmodified body shape. The animal's obstacle traversal was composed of four phases: approaching and collision, exploration, negotiation, and departure. (I) *Approaching and collision*. The animal ran rapidly (67 ± 11 cm $s^{-1}$) towards the beams in a normal, horizontal body orientation, collided with the beams, and was forced to stop. This occurred for all the trials during the escape response. (II) *Exploration*. In approximately one third of all trials (34 ± 20 %), before entering the obstacle field, the animal attempted to push through the beams (during which the body often pitched up slightly), then moved laterally (by at least one lateral beam separation, or 2 cm) to explore the first layer of beams in an attempt to find the boundary of the obstacles, during which it swept its antenna to detect openings. (III) *Obstacle negotiation*. The animal then negotiated the beam obstacles using one or a sequence of locomotor modes. (IV) *Departure*. In traversed trials, the animal ran away from the obstacle field and into the shelter, again, in a normal, horizontal body orientation.

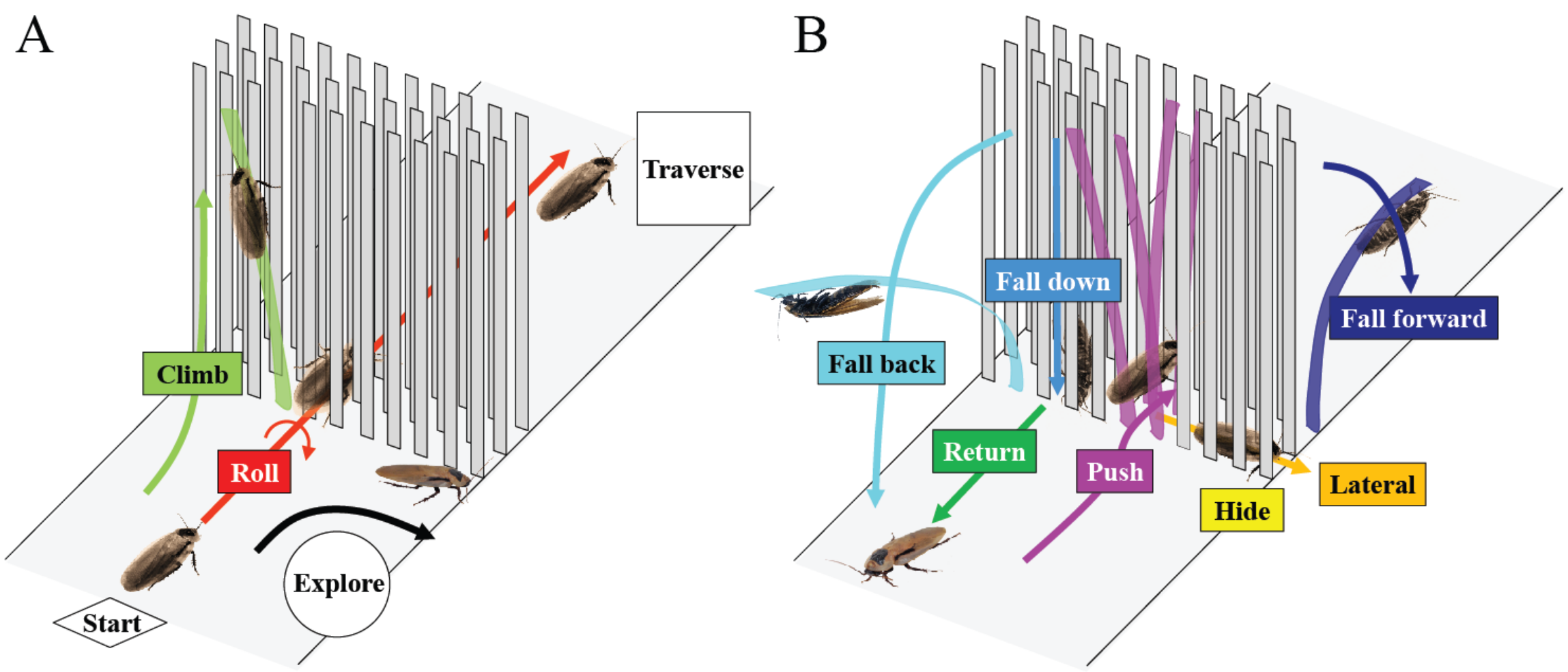

**Figure 5.** The discoid cockroach used a diversity of locomotor modes to negotiate beam obstacles. All trials began with and, if traversed, ended with running on flat ground in a horizontal body orientation. After a collision with the beams and possible exploration laterally along the first layer of beams, the animal used one or a sequence of locomotor modes to negotiate the beam obstacles. The dominant mode was a roll maneuver (red; movie S1, S2). Arrows indicate movement directions of each locomotor mode. Colors of locomotor modes match those in figure 6 and figures S1-S4. Colored beams indicate beams that experienced substantial bending due to animal interaction.

During the obstacle negotiation phase, the animal displayed a diversity of distinct locomotor modes (figure 5):

A. *Roll maneuver* (red; movies S1, S2): The animal rolled its body to either side (near 90º) such that its smallest body dimension (thickness) fit within the narrow lateral spacings between beams. In most cases after rolling, the animal then maintained its body orientation and maneuvered through the narrow gaps between the vertical beams by pushing its legs against the beams. In this mode, the legs adopted a more sprawled posture as compared to the running leg posture on flat ground and did not always use an alternating tripod gait. The pronotum joint also occasionally flexed and/or twisted slightly during the maneuver. Before the roll maneuver was initiated, the animal's body often briefly pitched up slightly as the animal pushed against the beams, but to a smaller degree than the body rolling. In some cases after rolling, the animal did not maneuver through the beams, but transitioned to other modes.
B. *Pushing* (magenta): The animal maintained a horizontal body orientation and pushed forward through the beam obstacles. In this mode, the animal's body was often elevated by the bent beams and its legs thus lost contact with the ground and pushed against the beams.
C. *Lateral movement* (orange): Between the layers of beams, the animal changed its heading turning left or right and moved in the lateral ($\pm y$) direction.
D. *Hiding* (yellow): The animal stopped moving and hid within the space between the beams of the obstacle field, likely using the shadows of the beams as a shelter.
E. *Returning* (light green): The animal returned (or occasionally stopped moving) without entering the beam obstacle field.
F. *Climbing* (dark green): The animal pitched its body head-up, grabbed onto the beams with its legs, and climbed upward along the beams.
G. *Falling forward* (cyan): After climbing up the beams, the animal fell to the ground across the beam obstacle field as the beams bent nearly 90º forward from the animal's weight.
H. *Falling back* (light blue): After climbing up the beams, the animal fell to the ground behind the beam obstacle field as the beams bent nearly 90º backward from the animal's weight.
I. *Falling down* (dark blue): After climbing up the beams, the animal lost foothold and fell down to the ground within the beam obstacle field.

Our locomotor pathway ethogram showed that movement of the animal through beam obstacles emerged *via* complex locomotor pathways (figure 6A). This differed from previous studies in insects and reptiles negotiating simpler obstacles like one or two steps, a gap, a ditch, or an incline (Watson et al. 2002a,b, Kohlsdorf and Biewener 2006, Theunissen and Durr 2013, Bernd et al 2002, Blaesing and Cruse 2004, Ritzmann et al 2005) in which only moderate changes in leg kinematics and body orientation were observed, and was most similar to cockroaches negotiating a shelf obstacle where more than one distinct locomotor mode (e.g., climbing and tunneling) was observed (Harley et al. 2009). This also contrasted with locomotion on simpler, near two-dimensional ground where the observed movements (e.g. running, walking, hopping, climbing) were often stereotyped.

### 3.2 High traversal performance dominated by roll maneuvers

Despite the considerable diversity of locomotor modes and pathways of traversal (figure 6A), we discovered that when body shape was unmodified, the animal most frequently (nearly half of the time) used roll maneuvers alone to traverse the beam obstacles (48 ± 9 %; table 4; figure 6A, red label; figure 8A, unmodified before, red), without employing any other mode(s). Traversal time using roll maneuvers alone (2.0 ± 0.5 s; figure 6B, red) was shorter than using any other locomotor mode(s) (5.6 ± 1.1 s; figure 6B, gray; $P < 0.05$, repeated-measures ANOVA). Closer examination of traversal time of trials using each of the other modes at least once confirmed that they all took longer than using roll maneuvers alone (figure 6B, colored bars; $P < 0.05$, repeated-measures ANOVA). The dominance of the fastest roll maneuvers accounted for the animal's high overall traversal performance (79 ± 12 % total traversal probability, table 4, figure 8A, unmodified before; 3.4 ± 0.7 s traversal time, table 4).

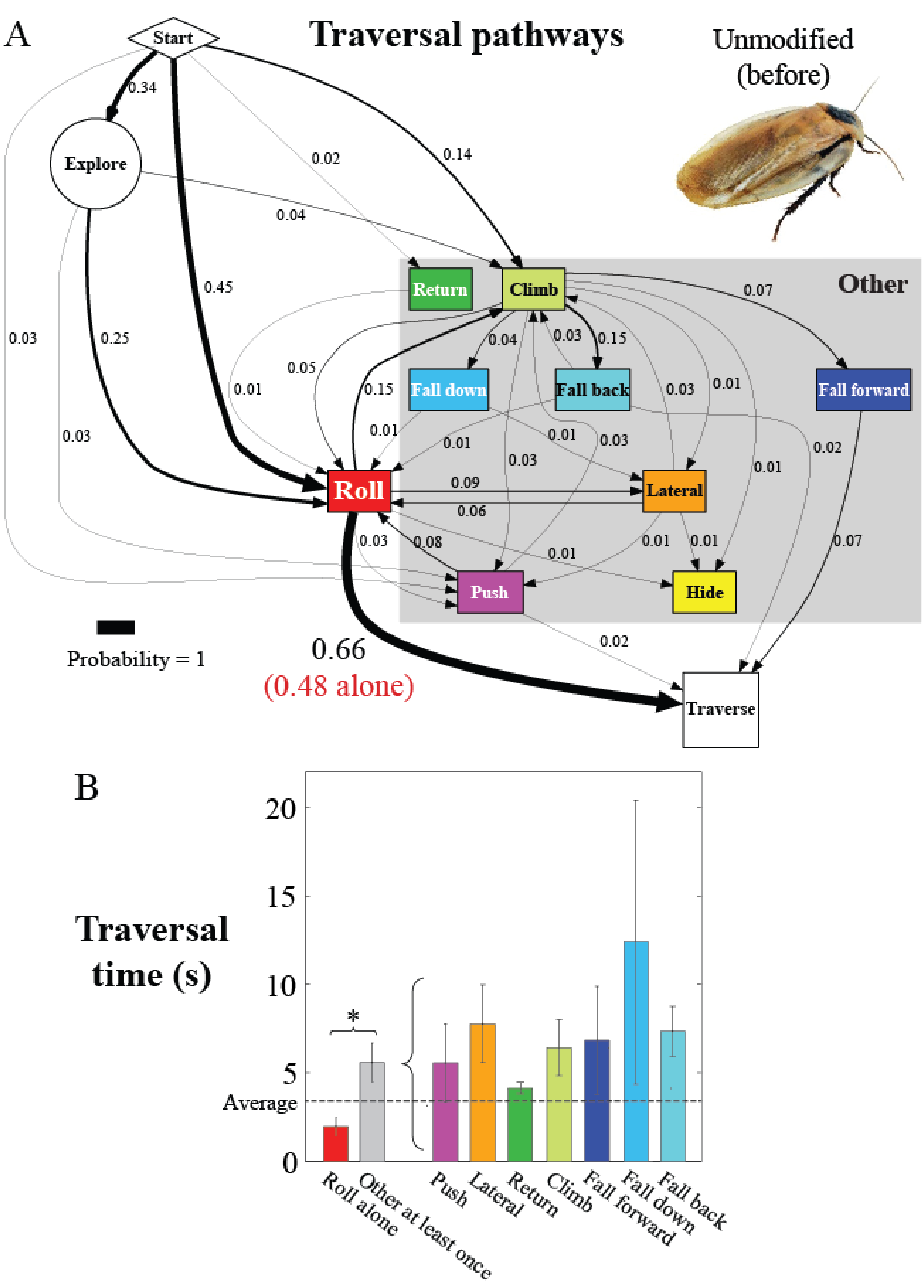

**Figure 6.** Cockroaches traversed beam obstacles using diverse locomotor pathways. With an unmodified body shape, traversal was dominated by the fastest roll maneuvers (movies S1, S2). (A) Locomotor pathway ethogram showing the animal's diverse locomotor pathways of traversal. Each colored box represents a distinct locomotor mode observed during obstacle negotiation. Arrows indicate transition between locomotor modes. The line width of each arrow is proportional to its probability, indicated by the number next to the arrow. Non-traversal probability is not shown for simplicity. Red label indicates traversal probability using roll maneuvers alone. (B) Comparison of traversal time between trials using roll maneuvers alone (red) and those in which other locomotor modes occurred at least once (gray). Colored bars are traversal times of trials in which each of the other locomotor modes was used at least once. Dashed line indicates average traversal time of all trials.

### 3.3 Body roundness increased traversal by facilitating roll maneuvers

#### 3.3.1 Animals locomoting with shells varying in shape

Artificial shells reducing body roundness altered the animal's beam obstacle traversal pathways and decreased traversal probability relative to unmodified body shapes (figures 6, 7; $P < 0.05$, repeated-measures ANOVA). Given the results from the unmodified shape, we focused on roll maneuvers and combined all other modes for easier comparison among treatments (see figures S1-S4 for locomotor pathway ethograms showing all locomotor modes). When the oval cone shell was added, the animal less frequently used roll maneuvers alone to traverse the obstacle field (22 ± 5 %), likely due to increased volume and surface friction (figure 7B, red label; table 4). Traversal using roll maneuvers alone was nearly inhibited with the flat oval (3 ± 4 %) and rectangle (2 ± 4 %) shells due to reduction in body roundness (figure 7C,D, red labels; figure 8A, red and white together; table 4). This decrease accounted for nearly the entire decrease in traversal probability, because traversal probability using any other mode at least once (figure 8A, gray) did not change significantly ($P > 0.05$, repeated-measures ANOVA).

Artificial shells reducing body roundness also slowed down the animal's traversal (figure 8B, table 4; $P < 0.05$, repeated-measures ANOVA). Traversal time for roll maneuvers alone (figure 8B, white) more than doubled with the oval cone shell (4.4 ± 1.2 s) due to increased volume and surface friction, and tripled with the flat oval and rectangle shells (6.5 s). Nevertheless, with each artificial shell, traversal using roll maneuver alone was always quicker than using any other mode(s) (figure 8B, gray).

Table 4. Animal beam obstacle traversal performance.

| | Condition | Unmodified (Before) | Oval cone | Flat oval | Flat rectangle | Unmodified (after) |
|---|---|---|---|---|---|---|
| Traversal probability (%) | All trials | 79 ± 12 | 50 ± 12 | 30 ± 16 | 22 ± 12 | 77 ± 12 |
| | Roll maneuvers alone | 48 ± 9 | 22 ± 5 | 3 ± 4 | 2 ± 4 | 47 ± 11 |
| | Other mode(s) at least once | 31 ± 8 | 28 ± 11 | 27 ± 15 | 20 ± 11 | 30 ± 5 |
| Traversal time (s) | All trials | 3.4 ± 0.7 | 8.0 ± 1.8 | 13.4 ± 6.9 | 13.6 ± 5.7 | 4.0 ± 0.5 |
| | Roll maneuvers alone | 2.0 ± 0.5 | 4.4 ± 1.2 | 6.5 ± 0.4 | 6.5 ± 0.0 | 2.8 ± 0.5 |
| | Other mode(s) at least once | 5.6 ± 1.1 | 11.0 ± 2.6 | 13.9 ± 6.8 | 14.2 ± 5.3 | 5.8 ± 1.0 |

All average data are given as mean ± s.d., unless otherwise specified.

Finally, the animal's obstacle traversal pathways, probability, and time all recovered upon removal of the artificial shells (figure 7E, figure 8, unmodified after, table 4; $P > 0.05$, repeated-measures ANOVA), confirming that these changes were not due to long-term learning or fatigue.

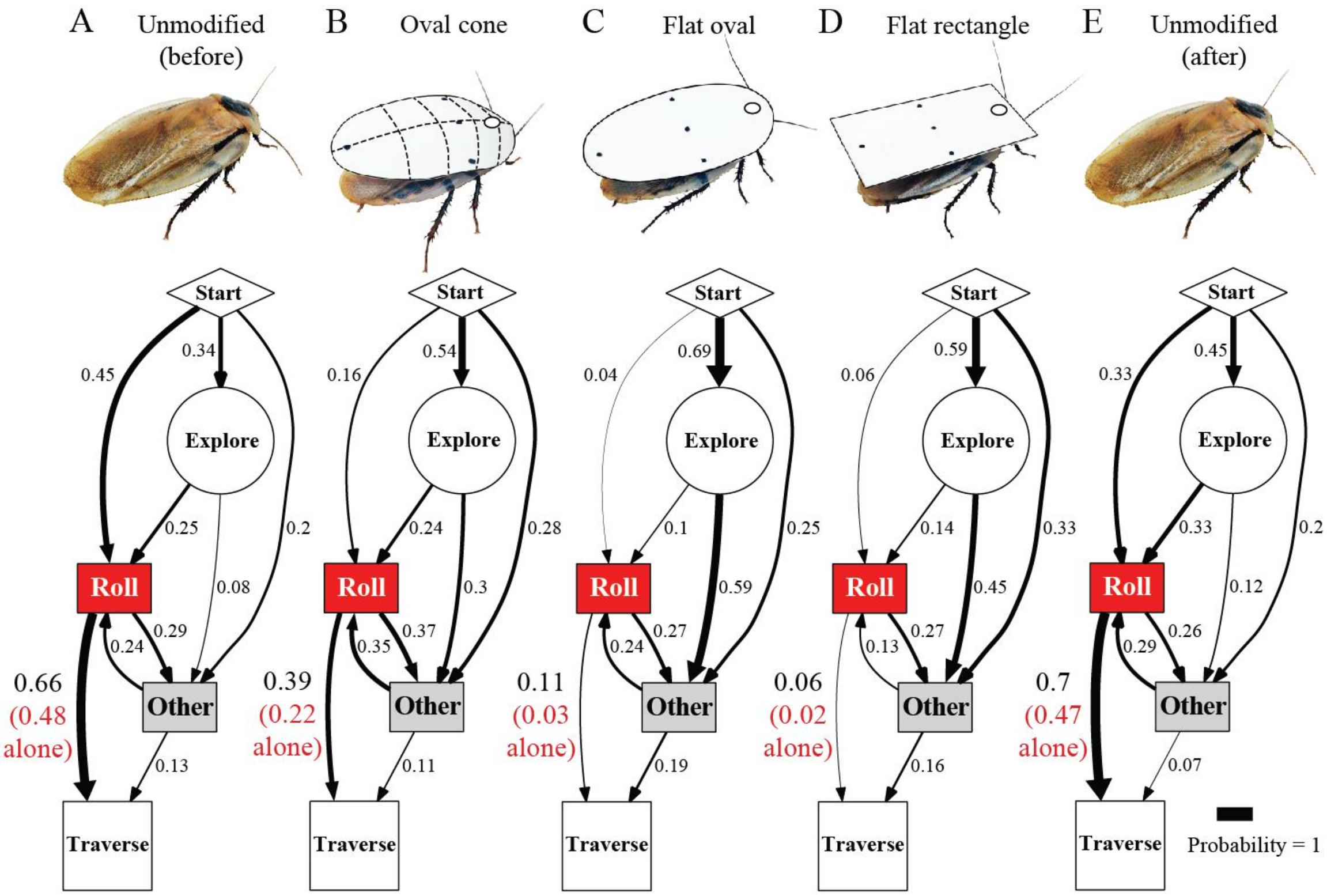

**Figure 7.** Cockroach's beam obstacle traversal pathways (A) were altered with artificial shells that reduced body roundness (B, C, D), primarily by inhibiting roll maneuvers (red box). Red labels indicate traversal probability using roll maneuvers alone. Obstacle traversal locomotor pathways

recovered upon removal of the shells (E). In the locomotor pathway ethogram, the line width of each arrow is proportional to its probability, indicated by the number next to the arrow. Red labels indicate traversal probability using roll maneuvers alone. Non-traversal probability is not shown for simplicity.

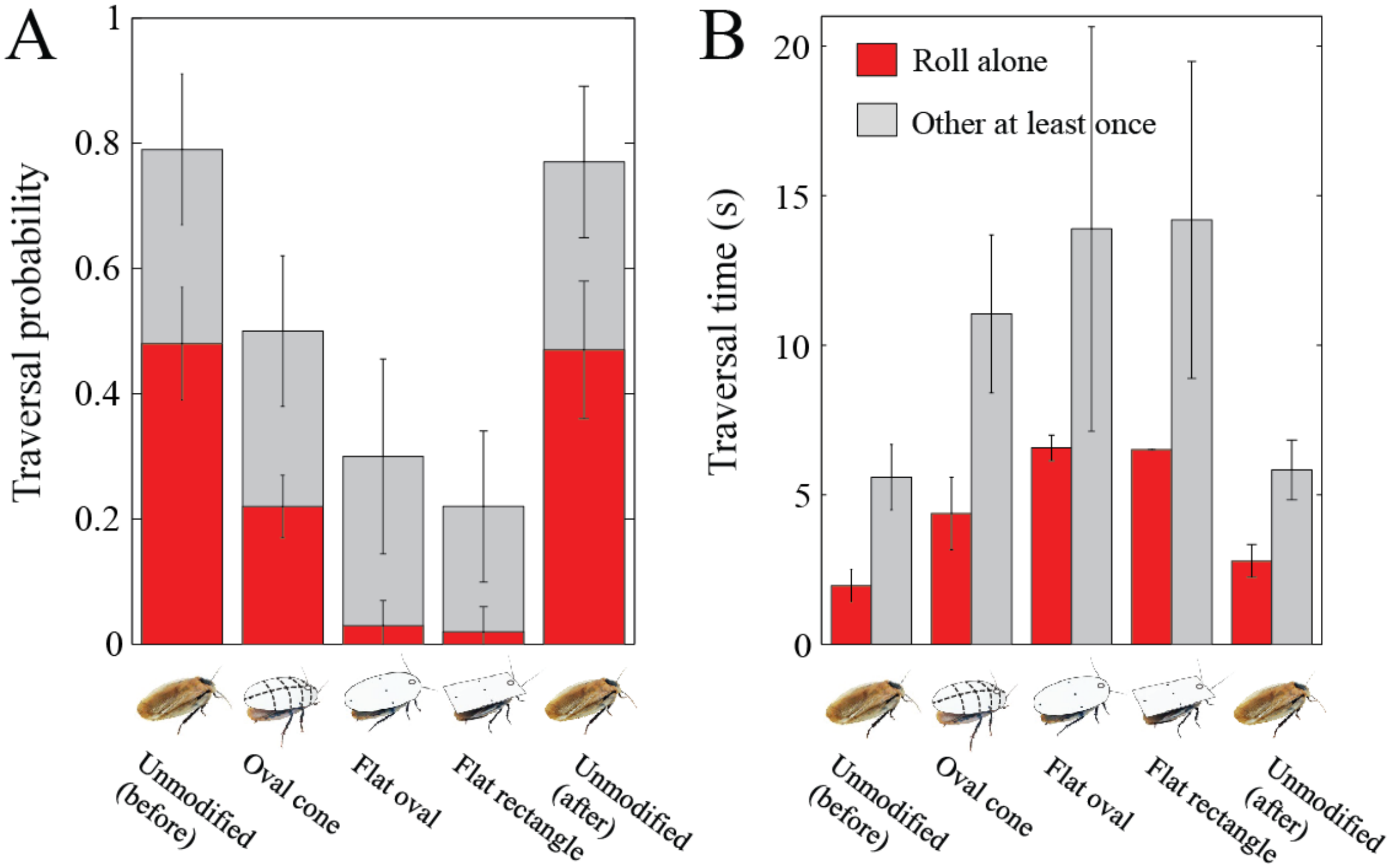

**Figure 8.** Cockroach beam obstacle traversal performance decreased with artificial shells that reduced body roundness, but recovered upon removal of the shells. (A) Traversal probability decreased with reduction of body roundness, primarily by inhibition of using roll maneuvers alone (red). (B) Traversal time increased with reduced body roundness, both for trials using roll maneuvers alone (red) and for trials using other locomotor mode(s) at least once (gray). Error bars are ±1 s.d.

3.3.2 Animals being pulled with shells varying in shape

Reduction of body roundness also hindered body rolling for deceased animals pulled through the beam obstacles, even when active leg thrusts and postural adjustments were not present. With unmodified shape and the oval cone shell, the animal body frequently displayed rolling as it was pulled through the beams (figure 9C). By contrast, with the flat oval and rectangle shells, the animal body almost always pushed forward through while maintaining a horizontal body posture (figure 9D).

Experiments on deceased animals suggested that the constant body vibrations from intermittent leg-ground contact and the ability of the body to pitch up to an appropriate angle appear important in living animals to induce body rolling in combination with rounded body shapes.

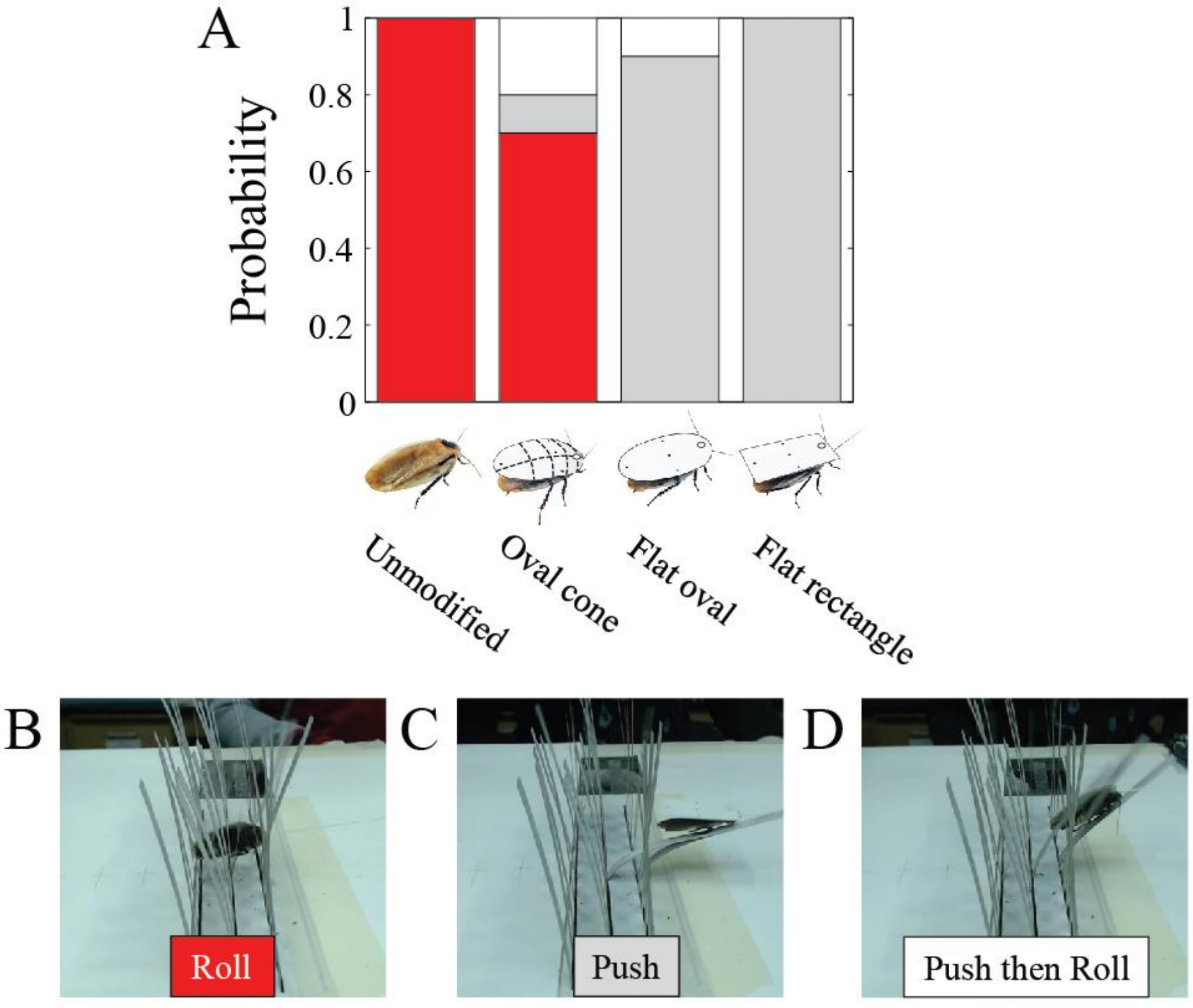

**Figure 9.** The motion of animal body being pulled through beam obstacle fields depended on body shape. (A) Probability distribution of animal's body motion during pulling at constant speed. Representative pictures of animal forward motion using (B) rolling, (C) pushing using a horizontal body posture, and (D) pushing followed by rolling.

Together, these observations demonstrated that, in addition to the performance loss due to increased volume and surface friction of the shells, reduced body roundness decreased the discoid cockroach's capacity to frequently and rapidly traverse beam obstacles by inhibiting the fastest roll maneuvers. This supported our hypothesis that a rounded body shape can enhance traversability through 3-D, multi-component obstacles such as grass-like beams.

### 3.4 Adding a cockroach-inspired rounded-shell enabled robot to traverse beam obstacles

Our discoveries of terradynamically streamlined shapes in cockroaches inspired a new approach that could enable robots to traverse densely cluttered obstacles rather than navigating around them. Traversing clutter terrain using shape is particularly useful for small robots (Birkmeyer et al 2009; Kovač et al 2010; Hoover et al 2010; Baisch et al 2011; Haldane et al 2015) whose size limits the deployment of onboard sensors, computers, and actuators for real-time obstacle sensing and path planning to cope with such terrain.

To test the shape design inspiration, we challenged our six-legged, open-loop robot, VelociRoACH, to traverse a beam obstacle field similar in configuration to, but larger than, that used in animal experiments (figure 3B). We found that with its unmodified, cuboidal body (figure 2B), the robot rarely traversed beams, even though its body (6.6 cm) was narrower than beam lateral spacing (10 cm). Instead, as soon as the body contacted a beam (resulting from constant body vibrations due to intermittent ground reaction forces), the robot turned to the left or right, but never rolled, and became stuck between adjacent beams either before entering or within the obstacle field (figure 10A; movies S1, S2). Only when the robot's trajectory was nearly straight and in the very middle of the two adjacent beams could it run through (figure S5) quickly (1.5 ± 1.8 s) without substantial contact (20% probability, figure 10C, unmodified).

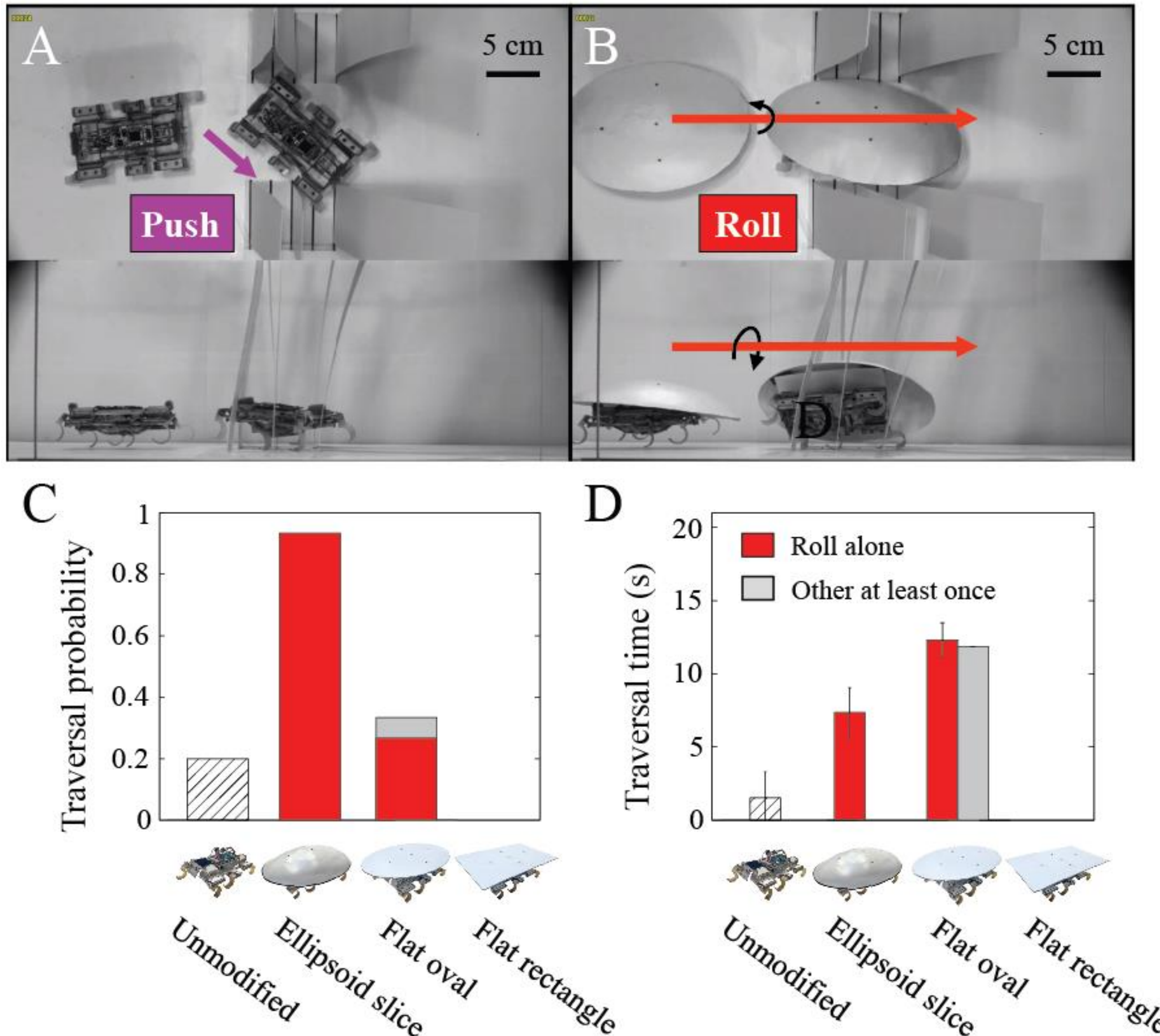

**Figure 10.** A rounded ellipsoidal shell enabled the robot to traverse beam obstacles. (A) Snapshot of the robot with an unmodified cuboidal body shape being stuck within the beam obstacles (movies S1, S2). (B) Snapshot of the robot with an ellipsoidal shell rolling its body to the side and maneuvering through the gap (movies S1, S2). The robot's (C) traversal probability and (D) traversal time showed similar dependence on shell shape as observed in the animal experiments. Hatched bars show the robot's traversal probability and traversal time with an unmodified cuboidal body narrower than beam lateral spacing. Error bars are ± 1 s.d.

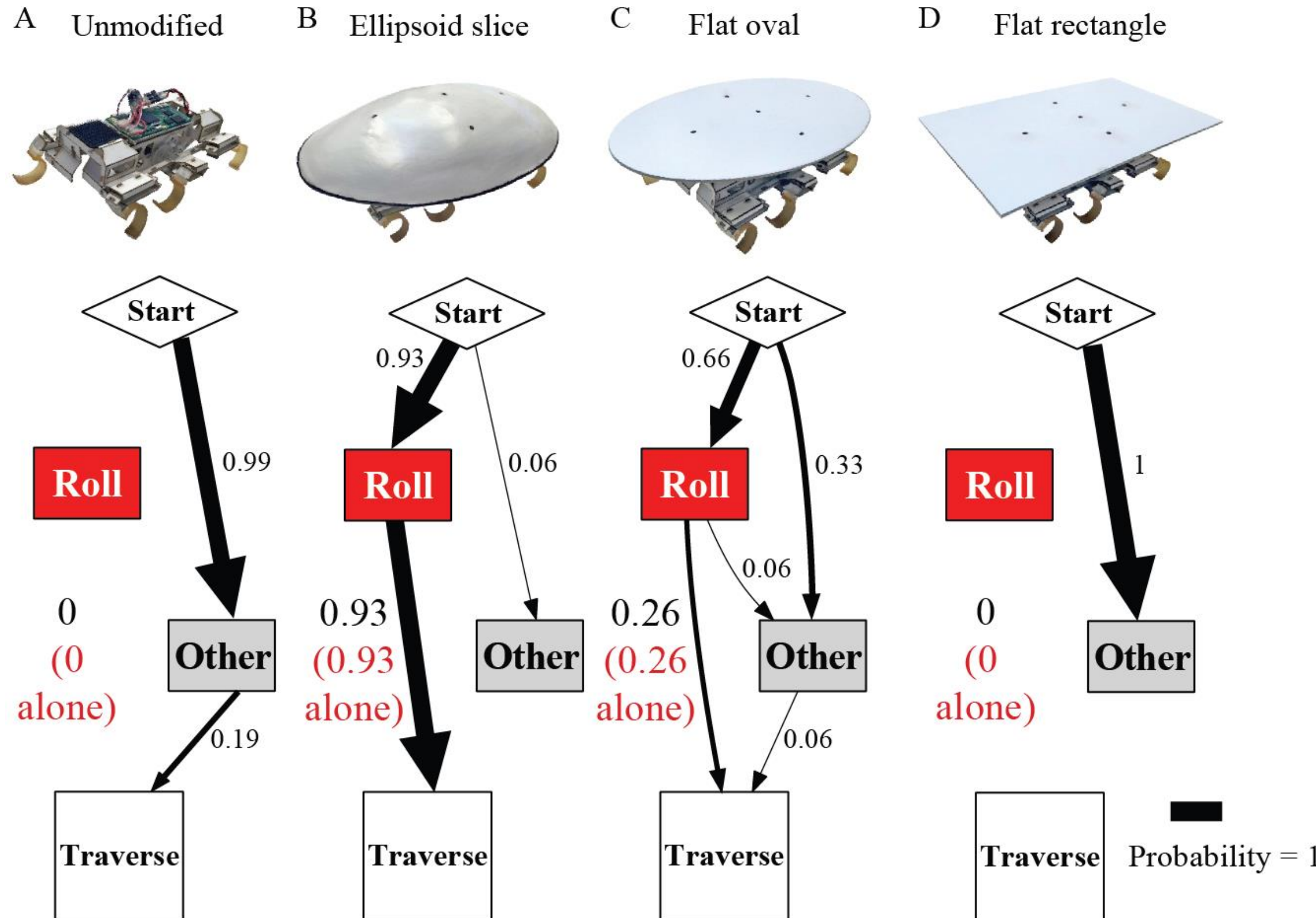

**Figure 11.** Robot's obstacle traversal pathways were altered by exoskeletal shells. In the locomotor pathway ethogram, the line width of each arrow is proportional to its probability, indicated by the number next to the arrow. Red label indicates traversal probability using roll maneuvers alone. Non-traversal probability is not shown for simplicity.

Upon adding a thin, rounded ellipsoidal shell inspired by our animal experiments, the robot traversed the beam obstacles with higher probability (93%; figure 10C, ellipsoid slice), even though the shell (12 cm) was wider than beam lateral spacing (10 cm). During traversal, the robot first pitched up while pushing against the beams, and then rolled to the side and maneuvered through the gap between the beams (figure 10B; movies S1, S2) rather than turning after contacting the beams. This was achieved using the same open-loop control without adding any sensory feedback. Further reduction of body roundness using the flat oval and rectangle shells resulted in reduction and eventually inhibition of roll maneuvers (figure 11), reducing traversal probability and increasing time (figure 10D), again similar to animal observations (figure 7). Together, these findings demonstrated that the mechanical interaction of the thin, rounded body with the beam obstacles alone was sufficient to induce roll maneuvers and facilitate beam obstacle traversal.

Besides facilitating body rolling *via* mechanical interaction with the beam obstacles, body roundness also assisted in maintaining the robot's heading. We observed that after the initial contact with the beams, the ellipsoidal and flat oval shells both more often drew the robot towards the middle of two adjacent beams, whereas the flat rectangle shell and the robot's unmodified cuboidal body more often resulted in the body turning away from one beam and towards the other, or even deflection of the robot laterally away from the beams (figures S5-S8).

We noted that the robot's beam obstacle traversal *via* roll maneuvers was not as effective as that of the animal. The animal traversed beam obstacles with lateral spacing (1 cm) less than half its body width (2.4 cm) and less than its standing height (1.2 cm). By contrast, the robot only traversed beam obstacles with lateral spacing (10 cm) about 80% its shell width (12 cm) and more than twice its standing height (3.8 cm). This is partly due to the robot's less flattened body aspect ratio (length: width: height = 10:6.6:3) compared to the animal (4.9:2.4:0.7) and its inability to adopt a more sprawled leg posture or slightly flex/twist the body. Furthermore, even with this moderately narrow obstacle gap, many parameters must be well tuned for the robot with the ellipsoidal shell to traverse by roll maneuvers, including the size of the shell relative to beam lateral spacing, the fore-aft position of the shell over the robot body, and the robot's stride frequency. Refinement of robot shell, body, and leg design could further improve its obstacle traversal performance using synergistic operation of body shape, body flexibility, leg morphology, and leg kinematics (Spagna et al. 2007).

### 3.5 Differences between animal and robot traversal modes

The robot did not show all the locomotor modes observed in the animal (figure 11). Only four modes resembled those of the animals: roll maneuver, pushing, climbing, and exploration. Moreover, these modes did not closely mirror the animal's corresponding modes (figure 5). (1) The robot often had to push against the beams for a substantial amount of time to pitch its body up substantially before it rolled to the side to maneuver through the beams (movies S1, S2). By contrast, while the animal's body also often pitched up before roll maneuvers, the duration of pitching was briefer (movies S1, S2). (2) During pushing, the robot's legs were always engaging the ground due to the only moderately cluttered beams, while the animal's body and legs were lifted off the ground due to bending of the more densely cluttered beams. (3) The robot's pushing often resulted in a lateral deflection away from the beams (figure 12, orange). The robot could not turn around and then try to negotiate the beams again like the animal could during exploration. (4) The robot's

"climbing" mode always resulted in a body pitch of nearly 90 degrees against a beam, but the robot was unable to lift off ground and actually climb up the beams, whereas the animals readily lifted themselves and climbed up onto the beams after substantial body pitch. (5) The robot more often flipped upside down (figure S5-S8) due to large pitch and roll vibrations from interaction with the beams, which was only very rarely (< 10 out of 869 trials) observed in the animal experiments. The animal could always quickly self-right, whereas the robot could never right itself.

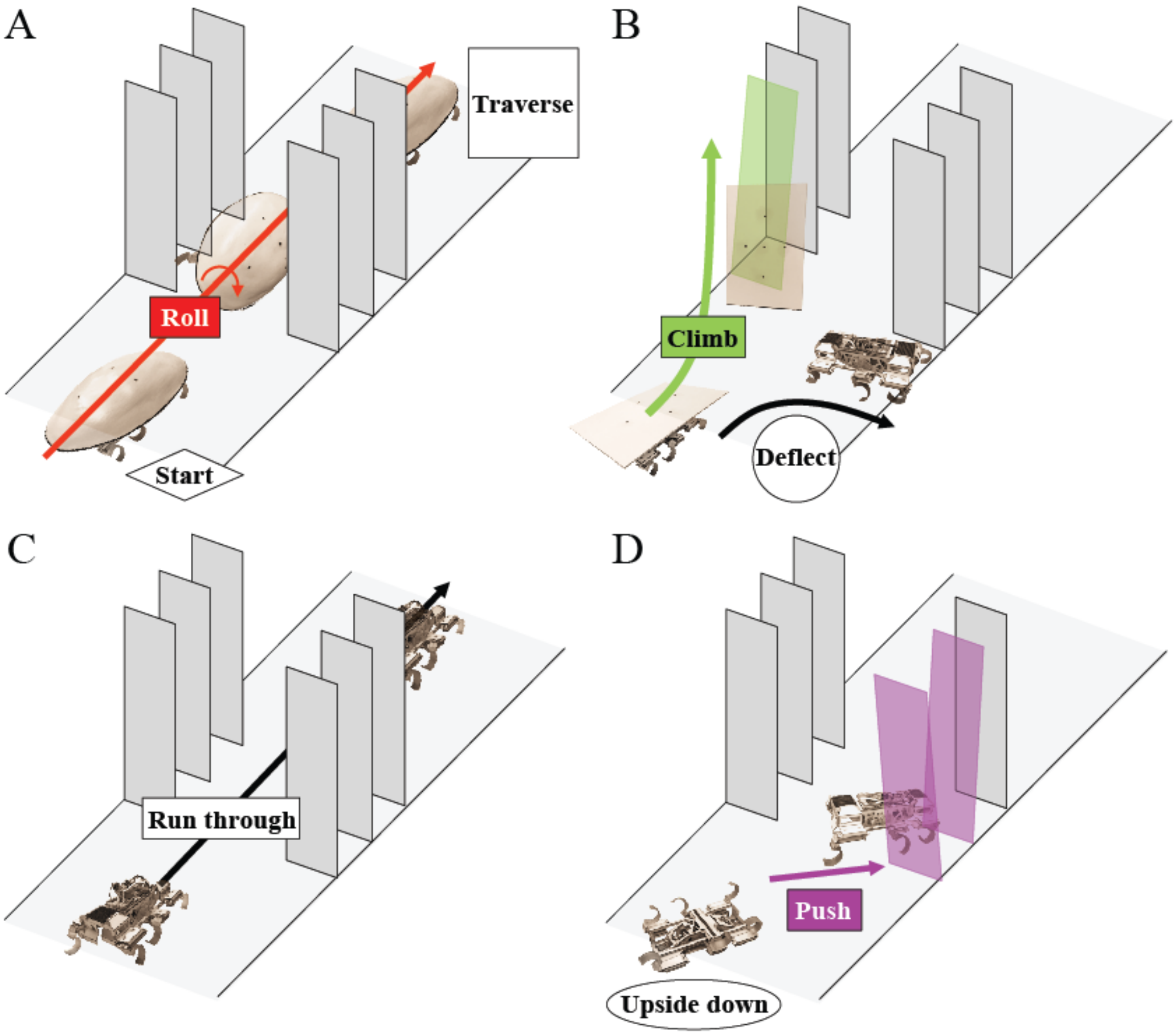

**Figure 12.** The robot's locomotor modes to traverse beam obstacles. The robot used one or a sequence of locomotor modes to negotiate the beam obstacles: (A) roll maneuver, (B) climbing and deflection, (C) running through without contact with the beams, and (D) pushing resulting in being stuck, and flipping upside down. Arrows indicate movement directions of each locomotor mode. Colors of locomotor modes match those in figures S5-S8. Colored beams indicate beams that were in contact with the robot during interaction.

Together, these differences in animal and robot obstacle traversal modes and pathways suggest that, besides rolling due to the mechanical interaction of a rounded body with the beams, the animal was likely employing sensory motor feedback to further assist traversal. The constant sweeping of antennae of the animal during exploration and beam negotiation phases (movies S1, S2) was likely to detect obstacle openings, guide path selection (Harley et al 2009, Zurek & Gilbert 2014), adjust leg posture gait, and flex/twist pronotum joint to actively facilitate body rolling to reduce terrain resistance. Recent studies of flight and swimming in analogous 3-D environments (Lin et al 2014, Flammang & Lauder 2013) discovered that birds and fish rely on visual and tactile sensory motor feedback to steer and navigate densely cluttered obstacles, rather than following pre-planned trajectories. Further animal experiments of rapid traversal in 3-D, multi-component terrain using electromyography (Sponberg & Full 2008) and manipulating the animal's sensors (Mongeau et al 2013, Harley et al. 2009), legs, and body flexibility will further elucidate the animal's capabilities and improve robot performance.

**3.6 Potential energy landscape model based on contact mechanics**

To begin to explain our observations of diverse locomotor pathways and the effect of locomotor shape, we developed a preliminary potential energy landscape model of locomotor-ground interaction during beam obstacle traversal (figure 13) inspired by recent success of potential field models for robot manipulation based on contact mechanics (Mason et al. 2011). In our minimal model, the *locomotor* (moving animal or robot) is simplified as a rigid body (ellipsoid) of weight *mg* with center of mass position (in the Cartesian lab frame) $(X, Y, Z)$ and orientation (Euler angles) $(\alpha, \beta, \gamma)$; two adjacent beams are simplified as two massless rigid plates with torsional springs (stiffness = $K$) at their bases (figure 13A). The total potential energy of the system is $E = mgZ + \frac{1}{2}K(\Delta\theta_1^2 + \Delta\theta_2^2)$, where $\Delta\theta_1$ and $\Delta\theta_2$ are the angular displacements of the two plates in the *x*-*z* plane due to contact with the locomotor. Considering the locomotor as it approaches the beams in the +*x* direction, $\Delta\theta_1$ and $\Delta\theta_2$ are uniquely determined by $(X, Y, Z, \alpha, \beta, \gamma)$, thus the total potential energy is a function of both translation and rotation of the body, i.e., $E = E(X, Y, Z, \alpha, \beta, \gamma)$. Here we focus on the initial contact phase where the bottom of the body touches the ground. Assuming that the body can rotate freely to access the lowest potential energy at any position in the horizontal plane, i.e., $E$ is minimized over the rotation degrees of freedom $(\alpha, \beta, \gamma)$, a potential energy landscape $E_0(X, Y)$ is obtained (figure 13B).

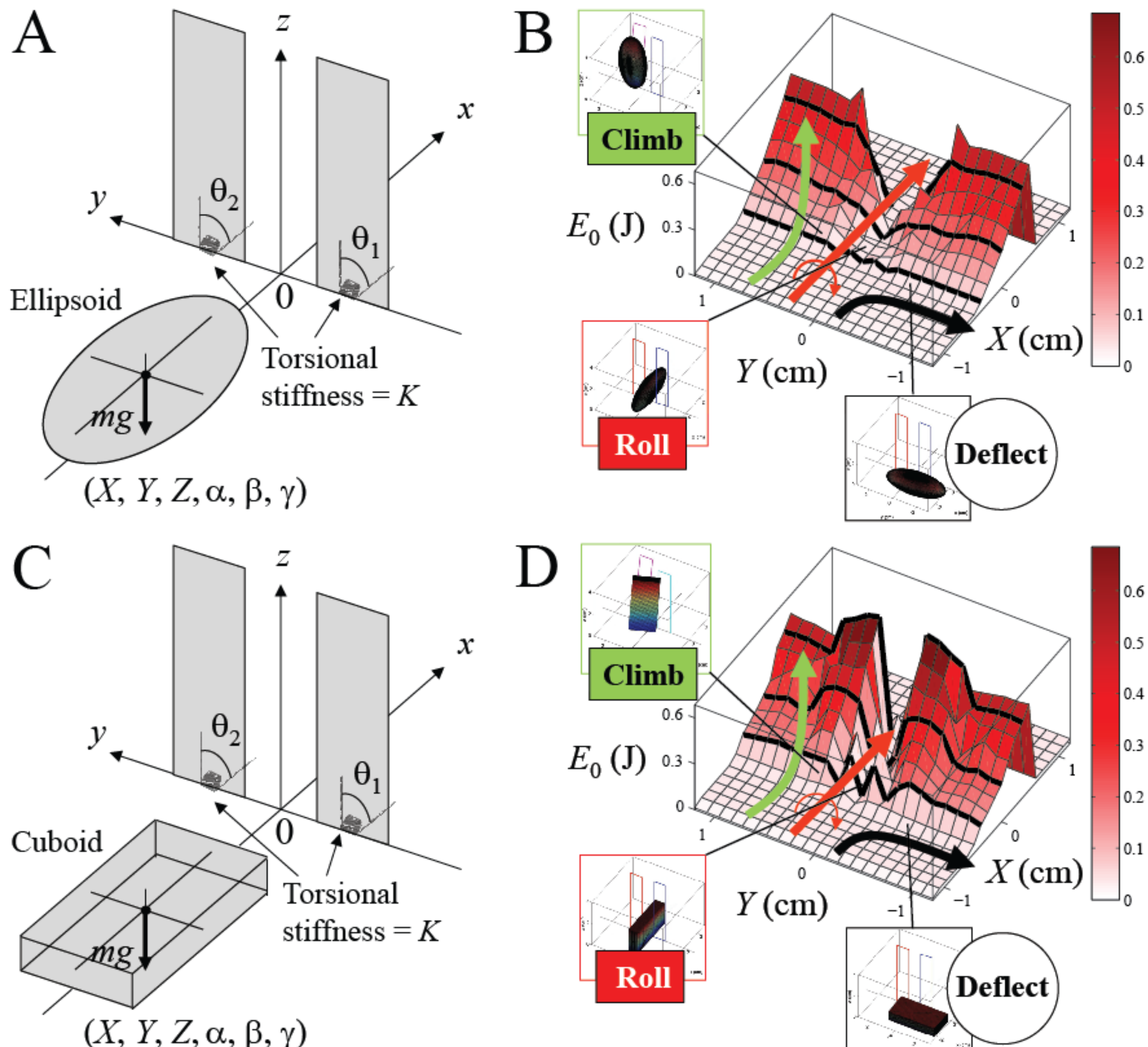

**Figure 13**. Potential energy landscape model of locomotor-ground interaction during beam obstacle traversal. (A) Simplification of the animal/robot locomotor as a rigid body (ellipsoid) and two adjacent beams as rigid plates with torsional springs at the bases. (B) Potential energy ($E_0$) landscape as a function of center-of-mass position ($X$, $Y$) of the ellipsoid in the horizontal plane. Insets in the green, red, and black boxes show the lowest potential energy orientations of the rigid body at three representative horizontal positions, which resemble the typical animal and robot body orientations observed in three locomotor modes shown by the green, red, and black arrows. (C, D) Potential energy landscape model for an angular locomotor body such as a cuboid. See text for definition of additional symbols.

The preliminary potential energy landscape model began to provide insights into the shape-dependent locomotor-ground interaction during movement in 3-D, multi-component terrain. From the potential energy landscape, interaction of the locomotor with two adjacent beams results in two high potential energy barriers around the locations of the beams, with a narrow and much lower potential energy barrier in between. The resulting lowest potential energy orientations of the locomotor are shown at three representative horizontal center-of-mass positions (figures 12B, insets). When the locomotor is exactly in between the two beams, its lowest potential energy orientation is to roll to its side. When it is close to one of the two beams, its lowest potential energy orientation is to pitch up. When it is close to one of the two beams, its lowest potential energy orientation is to yaw to the left or right. These orientations resemble the observed animal/robot body orientations during the beginning of a roll maneuver (red), climbing (light green), and exploration/deflection (gray) (figure 5A). We note that these lowest potential energy orientations are only instantaneous states of the locomotor, and do not fully represent the entire movement pathways (which comprise a series of orientation states). Further, the potential energy landscape is sensitive to locomotor shape. For example, compared to a rounded locomotor such as an ellipsoid (figure 13B), an angular locomotor (such as a cuboid with the same length, width, and thickness) results in two higher potential energy barriers with a narrower lower barrier in between (figure 13D). In addition, observation of the gradient of the potential energy landscape revealed that the angular cuboid results in repulsive lateral forces away from the lower barrier unless the locomotor is in the exact middle of the two beams (figure 13D, black curves), whereas the rounded ellipsoid results in attractive lateral forces towards the middle (figure 13B, black curves).

While beyond the scope of this study, we hypothesize that future developments of such locomotion energy landscapes based on contact mechanics may begin to allow statistical prediction of movement (e.g., probabilistic distribution of locomotor pathways) in 3-D, multi-component terrain and understanding terradynamic shapes. The theoretical framework of energy landscapes has been successfully used to understand the mechanisms of fundamental processes and/or making quantitative predictions in both close-to-equilibrium (e.g., protein folding, Wales 2003; DNA origami, Zhou et al 2015) and far-from-equilibrium and/or self-propelled systems (e.g., single molecule pulling experiments, Dudko et al 2006; cellular networks, Wang et al 2008; animal movement ecology, Shepard et al 2013, Wilson et al 2012). We observed that, analogous to statistical distribution of particles in energy states in an equilibrium system, animal and robot movement appeared to be more probable *via* locomotor pathways (e.g., roll maneuvers) that overcomes lower energy barriers (the lower energy barrier between the two high barriers; figure 13B), as suggested by our preliminary potential energy landscape models. We hypothesize that the constant body vibrations that the animal and robot experienced due to intermittent ground contact during legged locomotion may serve the function of "thermal fluctuations" to induce body rotation accessing locomotor pathways of lower potential energy barriers.

We note that our potential energy landscape model is only a minimal description of the locomotor-ground interaction physics and does not yet incorporate many possibly key elements, such as stochasticity, dynamics, driving forces, dissipation, and high-level, goal-directed behaviors of robots, and sensory feedback and even cognitive behaviors of animals. Future work is needed to more accurately measure contact physics and validate such potential energy landscape models, and complement them with dynamic simulation, stochastic modeling, models of driving and dissipative forces, and high-level behavioral models.

## Conclusions

By creating a new experimental apparatus to precisely control terrain parameters and modifying body shape in both animals and a robot as a physical model, we have discovered terradynamically "streamlined" shapes that enhance traversability in cluttered, 3-D, multi-component terrain. A thin, rounded body shape facilitates effective body reorientation to fit the smallest body dimension within obstacle gaps and reduce terrain resistance, analogous to the fusiform, streamlined shapes common in birds, fish, and aerial and aquatic vehicles that reduce fluid dynamic drag. While we have demonstrated body rolling facilitated by a thin, rounded shape in particular, body reorientation in other directions (e.g., yaw, pitch, or even a combination of roll, yaw, and pitch) facilitated by effective shapes interacting with the environment may also be useful in other forms of locomotion, such as running-climbing transition (Jayaram et al 2010), tunneling (Rollinson & Choset 2014), tobogganing (Wilson et al 1991), and self-righting (Domokos and Varkonyi 2008, Kovač et al 2010, Rabert 2012, Li et al 2015). For example, we observed that with the flat rectangle shell the robot was always attracted to one beam and pitched up (figure S8), which is useful for obstacle climbing.

Our quantitative results and early energy landscape model of how obstacle traversal depends on body shape also provide initial insights into the biomechanical mechanism of how terrestrial environments exert ecological forces on the evolution of the body shapes of animals living in them (Sharpe et al. in review). While our observations were made in vertical, grass-like beam obstacles of a given spacing, a thin, rounded body shape may have evolved in part to allow discoid cockroaches to rapidly and efficiently maneuver through clefts, slits, and crevices of various orientations and spacings (Jayaram et al 2013) common in their rainforest floor habitat. More broadly, thin, rounded body shapes found in many cockroaches and ground beetles may be an adaption for their life in dense grass and shrubs, loose bark and leaf litter, and boulders and rocks cracks (Bell et al 2007; Lovei & Sunderland 1996).

Our robot experiments demonstrated that, complementary to sensory feedback control and motion planning, challenging locomotor tasks such as obstacle negotiation could be accomplished by utilizing effective body shapes to interact with the environment without requiring sensors, computers, and actuators (Briod et al 2014) often impractical for small robotic platforms. This novel approach using distributed mechanical feedback and morphological computation (Pfeifer & Iida 2005) also offers robots unprecedented obstacle traversal capacity in highly cluttered terrain (i.e., obstacle spacing < robot size) common in natural and artificial environments, and can reduce time and energetic cost required for steering around obstacles altogether. Most existing obstacle avoidance approaches using high level feedback control (Leonard & Durrant-Whyte 1991, Thrun et al. 2000, Latombe 1996) are challenged by densely cluttered terrain (Koren & Borenstein 1991), largely because the locomotor-obstacle interaction models prevalent in these approaches such as artificial potential field (Khatib 1986, Koren & Borenstein 1989, Rimon & Koditschek 1992) and virtual force field (Borenstein & Koren 1989, 1991), are artificially defined to facilitate computation, but do not arise from contact mechanics essential in cluttered environments or capture important terrain topology and locomotor morphology.

Finally, we note that our study only provides a first demonstration of terradynamic shapes and is an initial step towards a principled understanding of biological and robotic movement in 3-D, multi-component terrain. Further parameter variation (Li et al. 2009) of three-dimensional, multi-component terrain and locomotor shape (e.g., Domokos and Varkonyi 2008) using novel controlled ground testbeds and robots as physical models (Li et al. 2009, Li et al. 2013), and development of new automated techniques to vary terrain and locomotor parameters (Qian et al. 2013) and quantify locomotor modes and pathways (e.g. Branson et al. 2009, Berman et al. 2014), will accelerate attaining this goal. We envision that creation of a broader and more general terradynamics (Li et al. 2013) describing the physical interaction between biological and robotic locomotors (and even manipulators, Mason et al. 2011) with their diverse, complex, three-dimensional terrestrial environments, will allow quantitative predictions of movement and terradynamic shapes. This advancement will not only broaden our understanding of the functional morphology, ecology, and evolution of terrestrial animals in their natural environments, but also guide the design of mobile robots that take advantage of terradynamic shapes to operate in the real world.

**Acknowledgements**

We thank Kaushik Jayaram, Tom Libby, Jean-Michel Mongeau, Nate Hunt, Dwight Springthorpe, Jaakko Karras, Max Donelan, Dan Goldman, Feifei Qian, Kipling Will, Olga Dudko, Milo Lin, and two anonymous reviewers for helpful discussions; Huajian Huang, Jeehyun Kim, and Michael Tsang for lab assistance; Fernando Bermudez for providing robot electronics; Tonia Hsieh for providing animals; the University of California, Berkeley Office of Laboratory Animal Care Facility for animal housing; and Armita Manafzadeh, Crystal Lee, Kristine Cueva, Will Roderick, and Mel Roderick for help with animal care. This work is funded by a Miller Research Fellowship to C.L. and Army Research Lab Micro Autonomous Systems and Technology Collaborative Technology Alliance to R.S.F. and R.J.F. Author contributions: C.L. R.S.F., and R.J.F. conceived study; C.L. designed study, performed animal and robot experiments, analyzed data, and developed model; A.O.P. and D.W.H. constructed and supported robot; H.K.L. performed deceased animal pulling experiments; and C.L., R.S.F., and R.J.F. wrote the paper.

**Supplemental Figures**

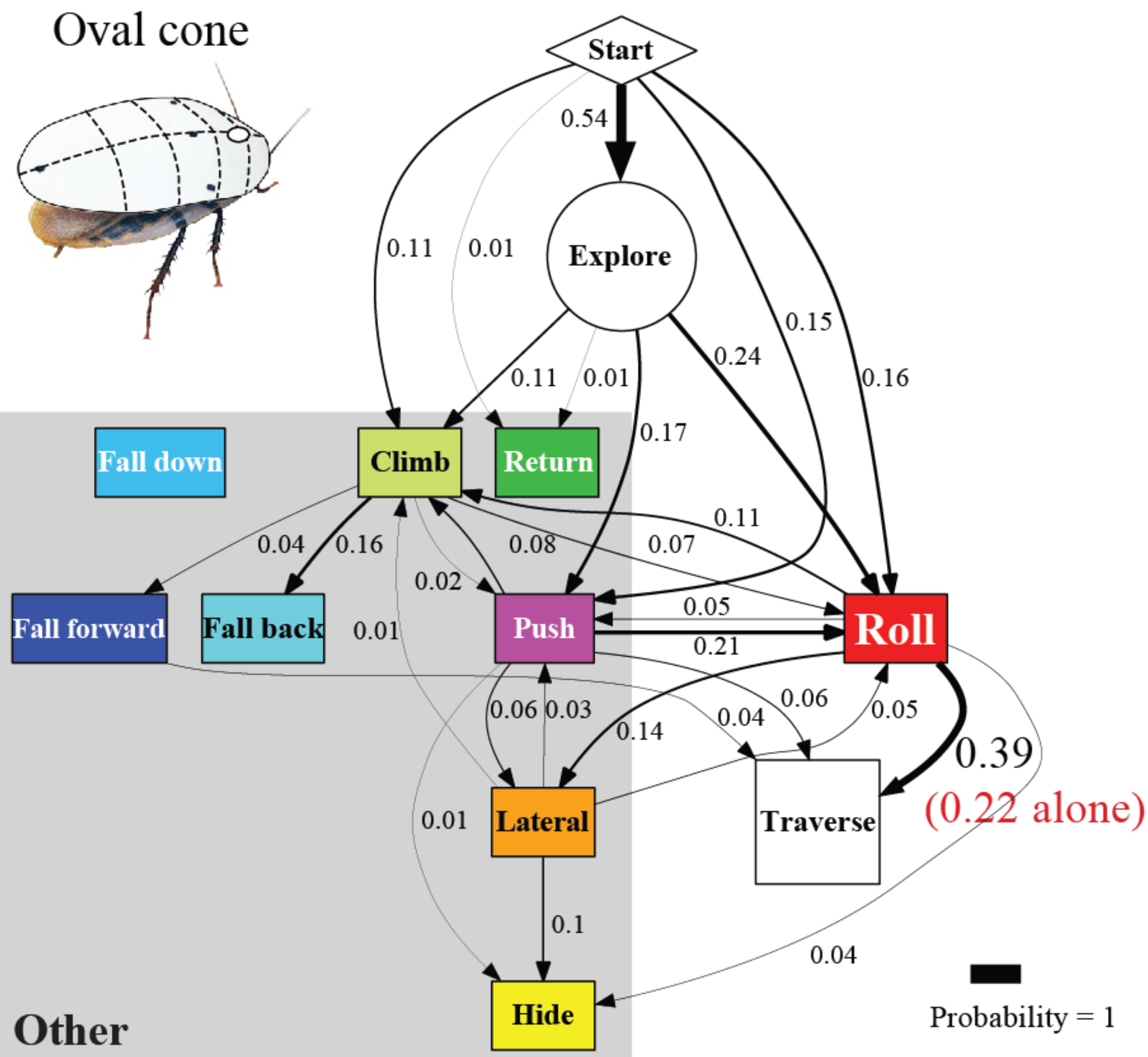

Figure S1. Cockroach's traversal process with the oval cone shell. In the locomotor pathway ethogram, the line width of each arrow is proportional to its probability, indicated by the number next to the arrow. Red label indicates traversal probability using roll maneuvers alone. Non-traversal probability is not shown for simplicity.

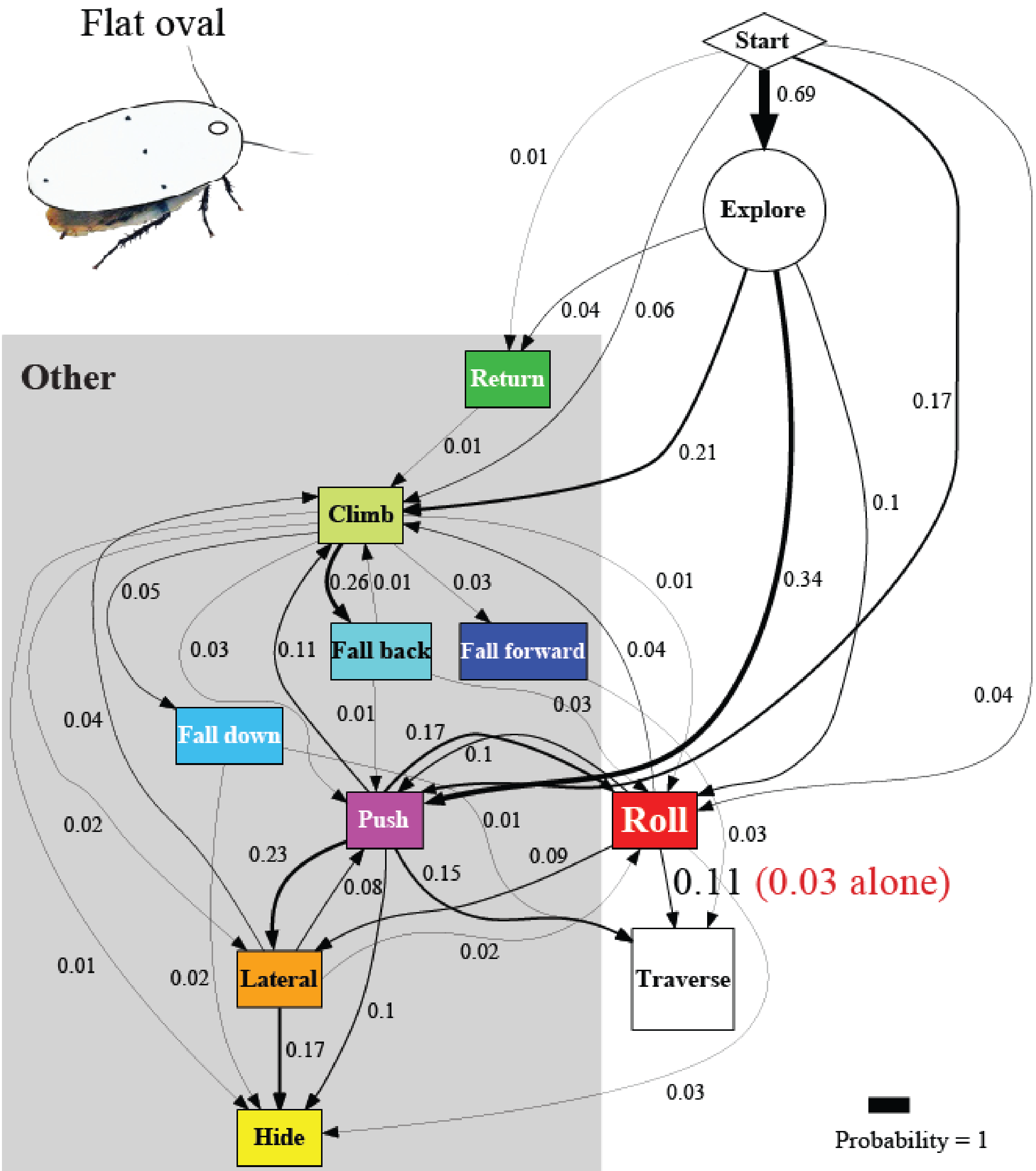

Figure S2. Cockroach's traversal process with the flat oval shell. $N$ = 5 animals, $n$ = 168 trials. In the locomotor pathway ethogram, the line width of each arrow is proportional to its probability, indicated by the number next to the arrow. Red label indicates traversal probability using roll maneuvers alone. Non-traversal probability is not shown for simplicity.

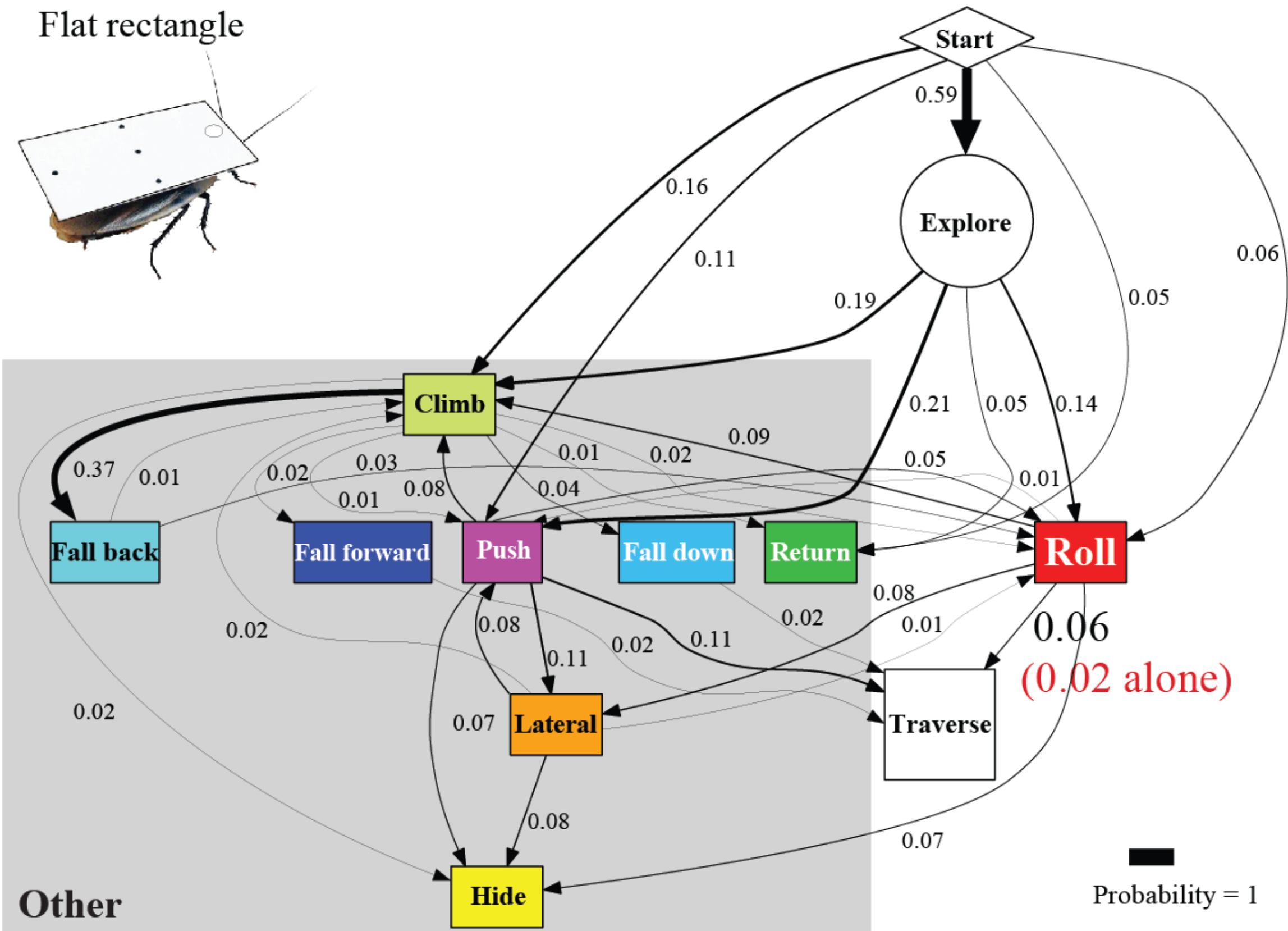

Figure S3. Cockroach's traversal process with the flat rectangle shell. $N = 5$ animals, $n = 204$ trials. In the locomotor pathway ethogram, the line width of each arrow is proportional to its probability, indicated by the number next to the arrow. Red label indicates traversal probability using roll maneuvers alone. Non-traversal probability is not shown for simplicity.

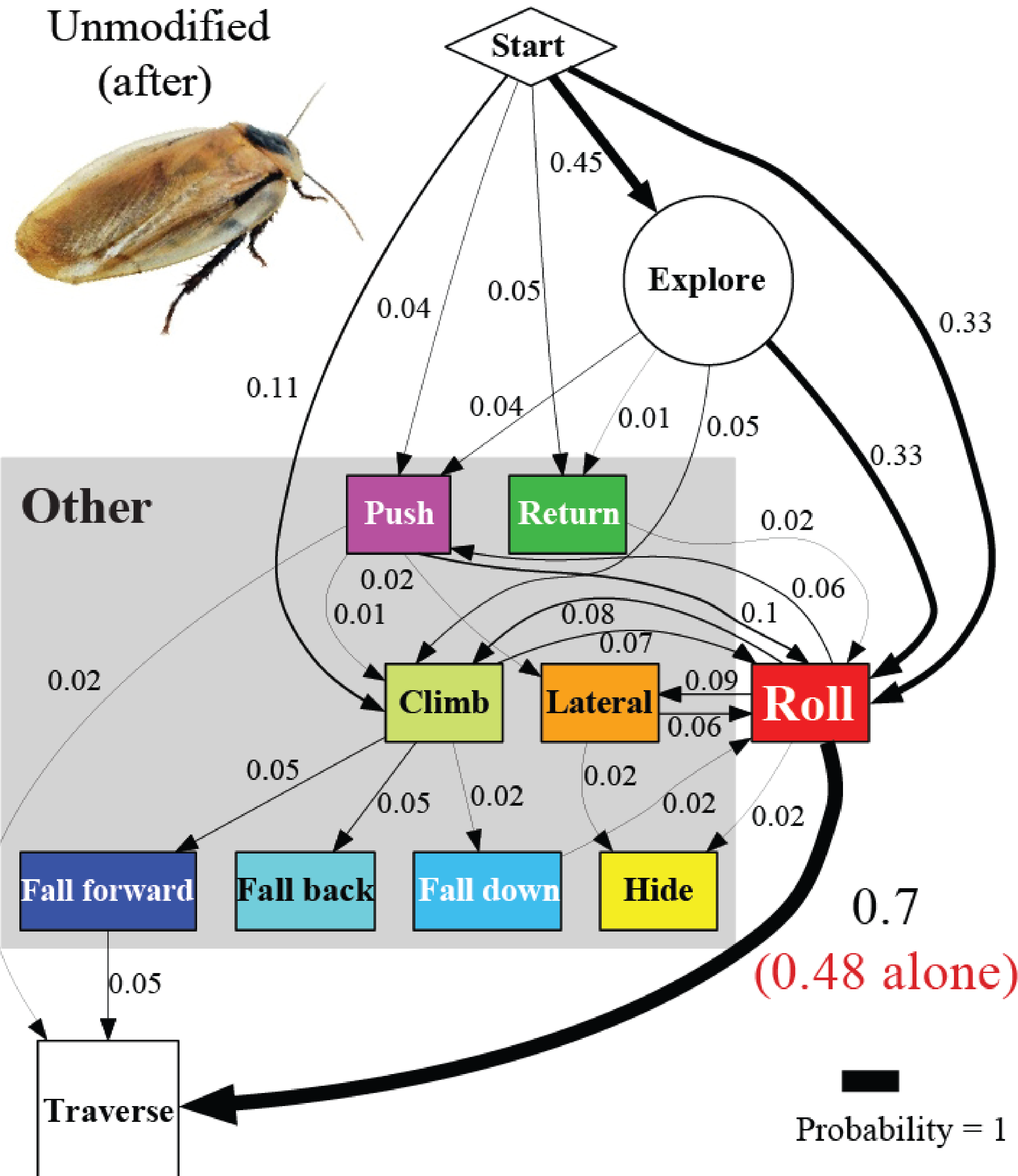

Figure S4. Cockroach's traversal process with an unmodified body shape after the shells were removed. $N$ = 5 animals, $n$ = 151 trials. In the locomotor pathway ethogram, the line width of each arrow is proportional to its probability, indicated by the number next to the arrow. Red label indicates traversal probability using roll maneuvers alone. Non-traversal probability is not shown for simplicity.

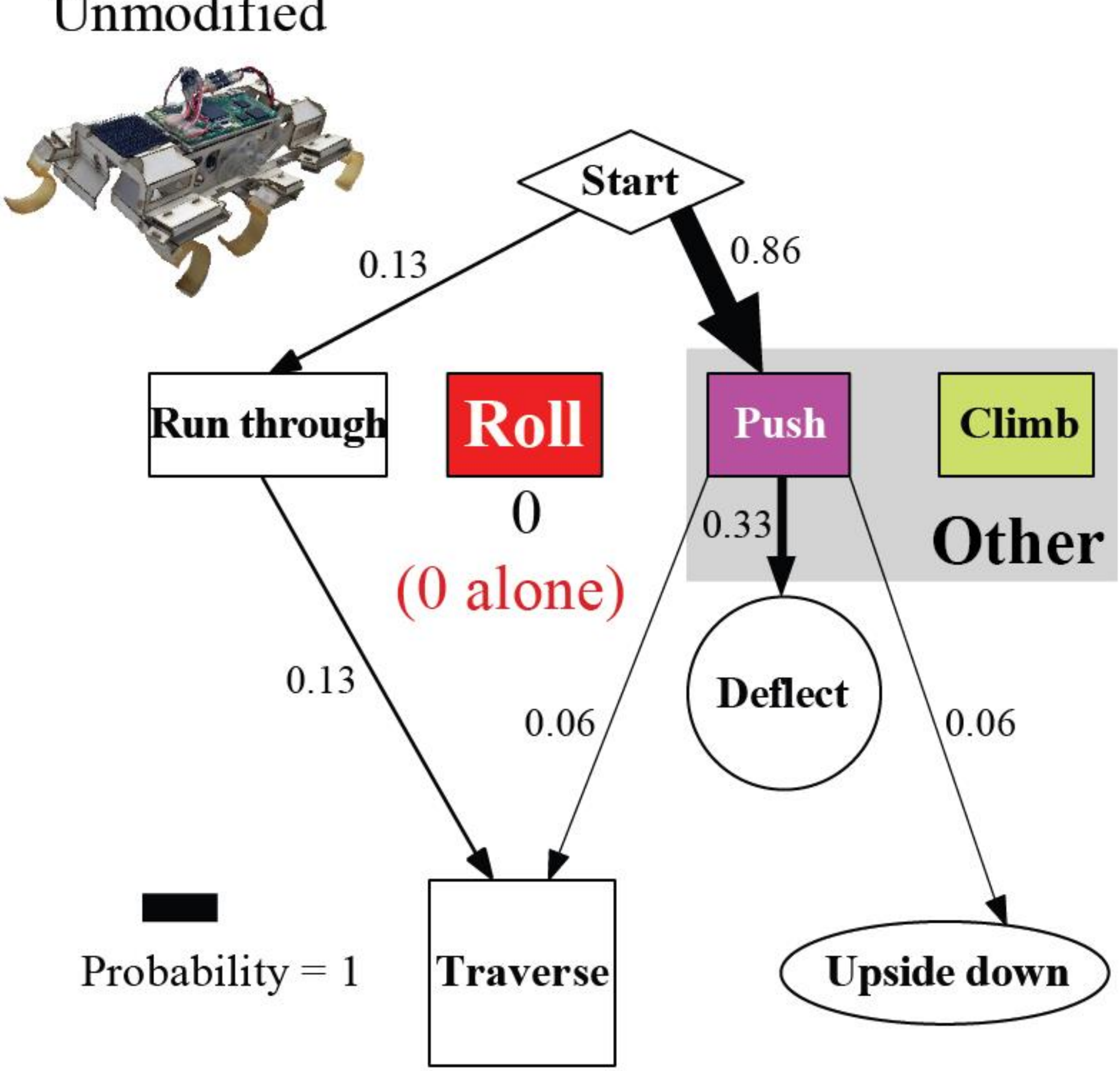

Figure S5. The robot's traversal process with an unmodified body shape. In the locomotor pathway ethogram, the line width of each arrow is proportional to its probability, indicated by the number next to the arrow. Red label indicates traversal probability using roll maneuvers alone. Non-traversal probability is not shown for simplicity.

Figure S6. The robot's traversal process with the ellipsoidal shell. In the locomotor pathway ethogram, the line width of each arrow is proportional to its probability, indicated by the number next to the arrow. Red label indicates traversal probability using roll maneuvers alone. Non-traversal probability is not shown for simplicity.

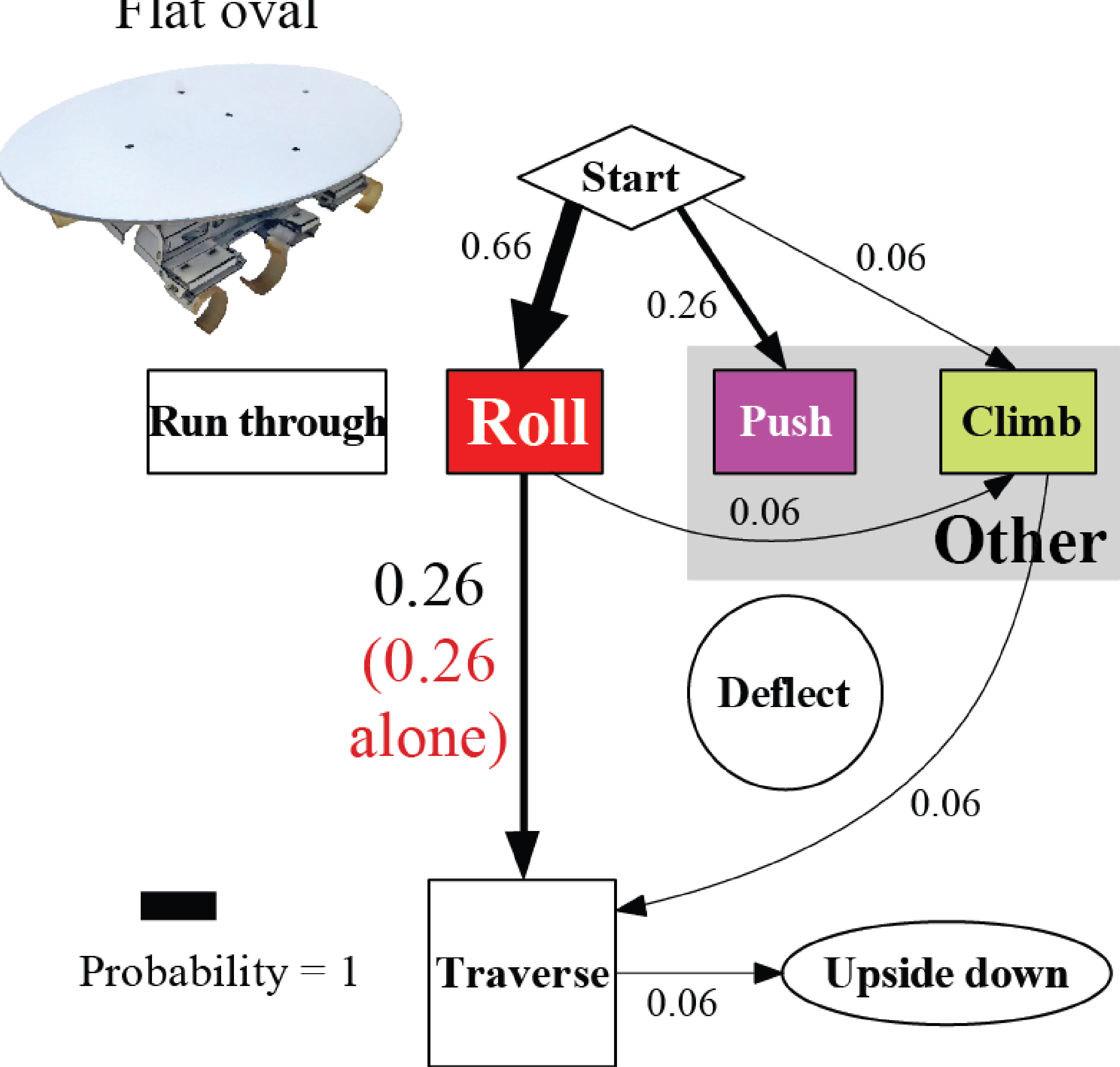

Figure S7. The robot's traversal process with the flat oval shell. In the locomotor pathway ethogram, the line width of each arrow is proportional to its probability, indicated by the number next to the arrow. Red label indicates traversal probability using roll maneuvers alone. Non-traversal probability is not shown for simplicity.

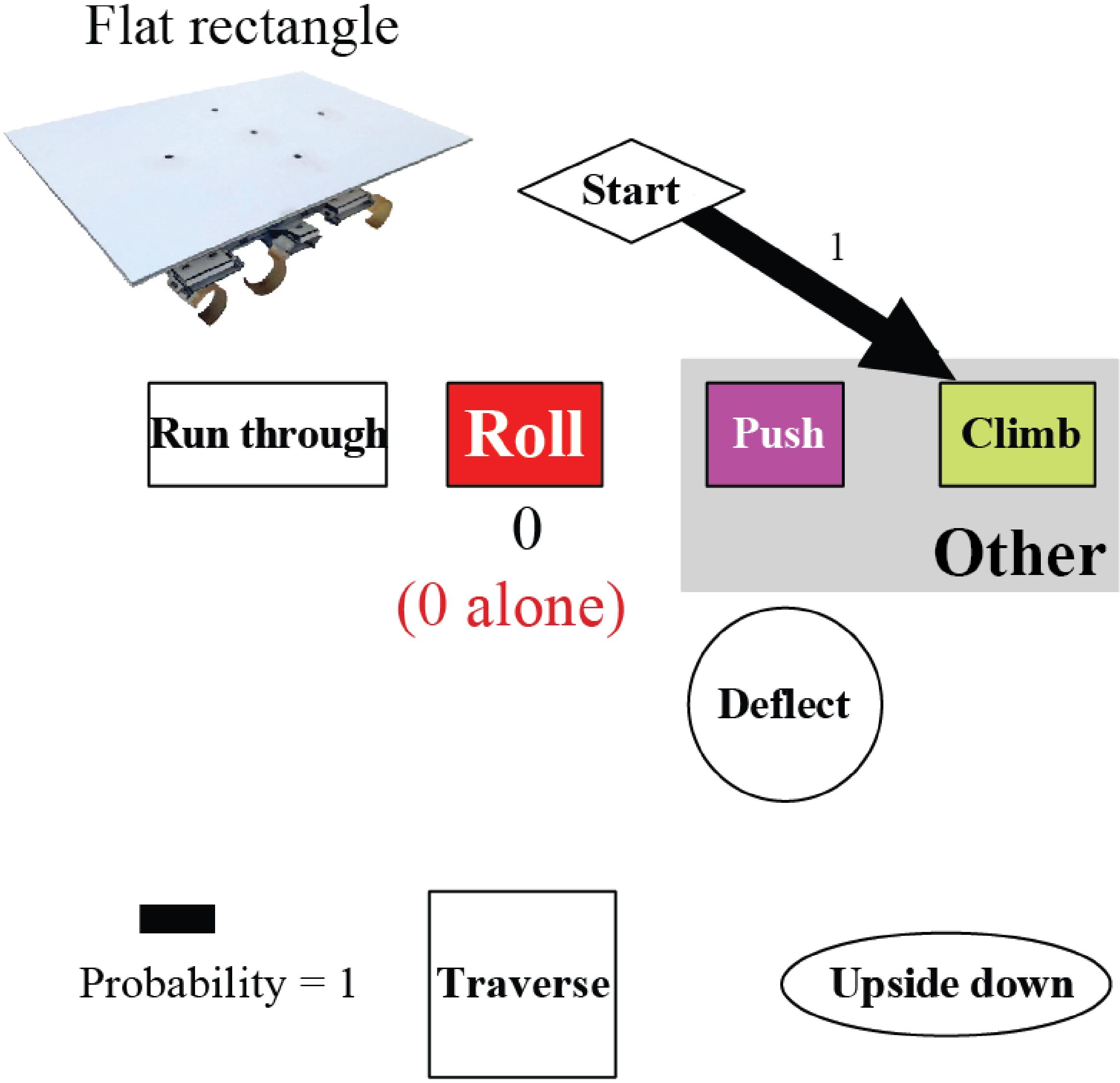

Figure S8. The robot's traversal process with the flat rectangle shell. In the locomotor pathway ethogram, the line width of each arrow is proportional to its probability, indicated by the number next to the arrow. Red label indicates traversal probability using roll maneuvers alone. Non-traversal probability is not shown for simplicity.

Movies S1

https://www.youtube.com/watch?time_continue=36&v=qZdGK0rcTAQ

Movie S2

https://www.youtube.com/watch?time_continue=64&v=PpZNw9Oxo3U